\documentclass[preprint2]{aastex}

\slugcomment{Not to appear in Nonlearned J., 45.}

\shorttitle{TeV $\gamma$-ray morphology of jets in radio galaxies}
\shortauthors{W. Bednarek}

\begin{document}


\title{Morphology of TeV $\gamma$-ray emission from the kpc scale jets in radio galaxies}


\author{W. Bednarek}
\affil{University of Lodz, Faculty of Physics and Applied Informatics, Department of Astrophysics, 90-236 Lodz, ul. Pomorska 149/153, Poland}

\email{bednar@astro.phys.uni.lodz.pl}



\begin{abstract}
TeV $\gamma$ rays are observed from a few nearby radio galaxies whose jets are viewed at relatively large angles towards the observer. This emission can be produced in the kpc scale jets whose Lorentz factors are decelerated from values of the order of several at the parsec scale distances. We consider in detail the model in which TeV $\gamma$-ray emission is produced by the relativistic electrons in the kpc scale jet which comptonize strongly beamed radiation from the inner (parsec scale) jet. As an example, we study morphology of the TeV 
$\gamma$-ray emission from the decelerated kpc scale jet in the nearby radio galaxy Cen~A. 
It is shown that TeV $\gamma$-ray emission can extend throughout the kpc scale distances, being relatively smoothly distributed along the jet for some parameters of the considered model. Investigation of the morphological structure of such specific $\gamma$-ray emission by the future Cherenkov Telescope Array (e.g. CTA) should provide important constraints on the content and dynamics of the kpc scale jet in Cen~A.
\end{abstract}




%
%
\section{Introduction}

Due to their proximity and large inclination angles of jets, radio galaxies represent a class of active galaxies 
in which the importance of the high energy processes in jets on kpc scale distances can be well studied. 
In fact, the closest object of this type, Cen~A, shows clear non-thermal emission extending up to X-ray energies (Feigelson et al.~1981). If interpreted as synchrotron emission, it requires the existence of electrons with TeV energies (e.g. Hardcastle et al. 2001). 
Note that such X-ray (and even soft $\gamma$-ray) emission can be also produced by the inverse Compton scattering of soft radiation produced in the central source (e.g. Skibo, Dermer \& Kinzer~1994, Brunetti, Setti \& Comastri~1997, Butuzova \& Pushkarev~2019).
In the case of Cen~A, the non-thermal X-ray emission from the jet is observed up to the distance of a few kpc (Kraft et al.~2002, Goodger et al.~2010, Tanada et al.~2019).
Electrons with TeV energies can naturally  produce $\gamma$-rays by scattering different types of soft radiation present within or around the kpc scale jet (e.g. Stawarz, Sikora \& Ostrowski~2003, Hardcastle \& Croston~2011, Wykes et al.~2015, Bednarek \& Banasi\'nski~2015, Bednarek 2019, Tanada et al.~2019). These electrons have to be accelerated locally in the jet due to their short cooling time scales (Hardcastle et al. 2003, Perlman \& Wilson~2005). 

The TeV $\gamma$-ray emission from the core region of Cen~A has been observed by the H.E.S.S. Collaboration (Aharonian et al.~2009). The spectrum is well described by a simple power law with a spectral index $2.52\pm 0.13_{\rm stat}\pm 0.20_{\rm sys}$ in the energy range 0.25$-$6 TeV (Abdalla et al.~2018). This spectrum smoothly connects to the hard GeV $\gamma$-ray component observed by the {\it Fermi} satellite (Abdo et al.~2010, Sahakyan et al.~2013, Brown et al.~2017, Sahakyan et al 2018). At least a part of the TeV $\gamma$-ray emission has been localized to be produced within the kpc scale jet in Cen~A (Sanchez et al.~2018), confirming the essential role of relativistic particles also in the kpc scale jets of radio galaxies. At present, 42 radio galaxies are known to emit $\gamma$ rays in the GeV energy range (The Fourth Fermi source catalog, Abdollahi, S. et al.~2019). It is not clear at present what is the production site of this $\gamma$-ray emission, either the inner jet or the large scale jet, or a combination of both. 

Recently, we proposed that TeV $\gamma$-ray emission from kpc scale jets of radio galaxies can be efficiently produced by electrons which inverse Compton (IC) up-scatter strongly boosted soft radiation from the parsec scale jets (Bednarek 2019). In fact, this type of soft radiation can dominate in the kpc scale jet over other types of radiation fields, provided that the jet is already significantly decelerated from pc to kpc scale distances, e.g. due to the entrainment of matter from stars into the jet or from the surrounding medium. In fact, the jet in Cen~A is observed to be subluminal on the hundred parsec distance scale, with the apparent speed $\sim 0.5$c (Hardcastle at el.~2003). 
According to the unification model of active galactic nuclei (AGNs), the inner jets in radio galaxies are expected to move with large Lorentz factors (of the order of several), as observed in blazars which are viewed at small observation angles. The jet in Cen~A is expected to be oriented at a large angle to the line of sight, estimated in the range $\theta\sim (12-45)^\circ$ (M\"uller et al.~2014) and  $\sim (50-80)^\circ$ (Tingay, Preston \& Jaucey~2001).

Motivated by the reports on the TeV $\gamma$-ray emission from the kpc scale jet in Cen~A, we investigate the morphology of such energetic emission, applying specific model for the velocity structure of the jet during its deceleration on the kpc distance scales. We predict the TeV $\gamma$-ray fluxes and spectra from specific regions within the kpc scale jet as a function of the Lorentz factor of the inner jet, its viewing angle, magnetic field strength, and the parameters describing the spectrum of relativistic electrons. Future comparison of the model predictions with detailed observations of the TeV $\gamma$-ray emission from the jet in Cen~A (by the HESS and the CTA) should allow to constrain the content of the jet on the kpc scale distances, the acceleration process of electrons to TeV energies, and also put light on deceleration mechanism of jets in AGNs.   

We do not intend to propose a complete description of the high energy emission from the jets of active galaxies  at different distance scales (starting from sub-parsec and finishing on hundreds of kpc) since the jet basic features look quite different on different distance scales (e.g. variability of emission, content of the jet, surrounding medium, etc). It is not even clear whether unique radiation and acceleration mechanisms can operate in the jet at different distances from its base. 
A few general models have been proposed, soon after detection of $\gamma$-ray emission from active galaxies early in 90-ties, in order to explain strongly variable high energy processes in the parsec scale jets (e.g. Ghiselini et al. 1992, Sikora et al.~1994, Mannheim \& Biermann 1992). Many modifications of these  models have been studied more recently but they do not predict directly emission features from the jet at the kpc scale distances.
Therefore, there is a need to investigate specific models that are able to explain emission features at much larger distances from the jet base (some of those models already mentioned above).

\begin{figure*}
\vskip 6.truecm
\includegraphics{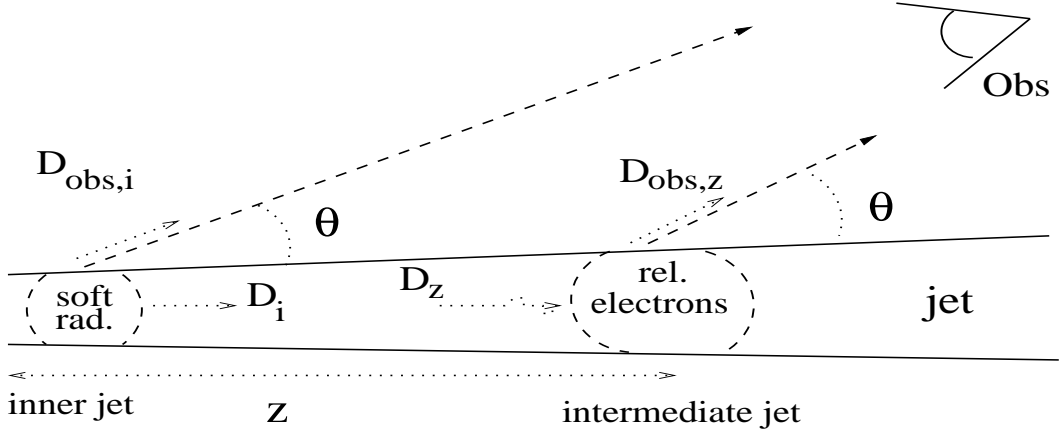}
\caption{Schematic representation of the jet in a radio galaxy. The inner, parsec scale, jet moves relativistically, with the Lorentz factor $\gamma_{\rm i}$ which is of the order of several. Relativistic electrons produce soft radiation towards the observer located at the angle $\theta$. This radiation is boosted towards the observer with the Doppler factor $D_{\rm obs, i}$. The intermediate jet, at the distance $\sim$0.1-10 kpc from its base, starts to be significantly decelerated, e.g. due to the entrainment of background matter. The outer part of this jet, at the distance $z$ from its base, moves with the Lorentz factor $\gamma_{\rm z}$. The soft radiation (from the inner jet) is relativistically boosted along the jet with the Doppler factor $D_{\rm i}$. It is de-boosted in the reference frame of the intermediate jet with the Doppler factor $D_{\rm z}$. The TeV $\gamma$ rays, produced in the kpc jet, are again boosted towards the observer with the Doppler factor $D_{\rm obs,z}$.}
\label{fig1}
\end{figure*}
\section{Comptonization of the inner jet radiation}

We investigate details of the scenario for the TeV $\gamma$-ray production in radio galaxies recently discussed by Bednarek (2019). Our aim is to predict the geometrical structure of the $\gamma$-ray emission from the kpc scale jet that might be observed with the future Cherenkov telescopes.
In the considered model, $\gamma$-rays are produced by relativistic electrons in the mildly relativistic (or sub-relativistic) kpc scale jet. They comptonize collimated soft radiation produced in the relativistic parsec scale jet (for details see Fig.~1). For the example purposes, we apply the soft radiation field which is assumed to be produced in the inner part of the jet in the radio galaxy Cen~A. The differential flux, $F_{\rm obs}$ (ph.~MeV$^{-1}$~cm$^{-2}$~s$^{-1}$), of this radiation is analytically approximated by Eq.~1 in Bednarek (2019), based on the observations by Mirabel et al.~(1999) and Marconi et al.~(2000). The observed soft radiation from Cen~A is assumed to be relativistically enhanced in the reference frame of the kpc scale jet due to the relativistic boosting effect. The external observer is located at the angle $\theta$ in respect to the jet axis. The inner jet moves with the large Lorentz factor, $\gamma_{\rm i}$, and the velocity, $\beta_{\rm i} = v_{\rm i}/c$, normalized to the velocity of light $c$. Relativistic electrons are located in the region which is at the distance $z$ from the base of the jet. This part of the jet moves with the mild Lorentz factor, $\gamma_{\rm z}$, and the velocity 
$\beta_{\rm z} = v_{\rm z}/c$. Then, the density of soft radiation field in the reference frame of the kpc scale jet (in which the relativistic electrons are injected) is calculated from,
\begin{eqnarray}
n(\varepsilon') = F_{\rm obs}(\varepsilon) 
{{d_{\rm L}^2}\over{z^2c}}{{D_{\rm i}^2}\over{D_{\rm obs,i}^2}}D_{\rm z}^2, 
\label{eq1}
\end{eqnarray}
\noindent
where $d_{\rm L} = 3.8$ Mpc is the luminosity distance to Cen~A (Harris, Rejkuba \& Harris~2010),  $D_{\rm i} = [\gamma_{\rm i}(1 - \beta_{\rm i})]^{-1}$ is the Doppler factor of the inner jet as observed along the jet axis, $D_{\rm obs,i} = [\gamma_{\rm i}(1 - \beta_{\rm i}\cos\theta)]^{-1}$ is the Doppler factor of the inner jet at the direction towards the observer, $D_{\rm z} = [\gamma_{\rm z}(1 + \beta_{\rm z})]^{-1}$ is the Doppler factor of the radiation field approaching the kpc scale jet from the direction of the inner jet,  
$\varepsilon = \varepsilon' D_{\rm obs,i}/(D_{\rm i}D_{\rm z})$ is the soft photon energy in the observer's reference frame, and $\varepsilon'$ is its energy in the reference frame of the kpc scale jet.

We assume that the angular distribution of radiation field originating in the inner jet is almost mono-directional for the relativistic electrons in the reference frame of the kpc scale jet. This is a good approximation provided that the parsec scale jet is highly relativistic and the opening angle of this part of the jet is already relatively low. On the other hand, the relativistic electrons are assumed to have isotropic distribution in the reference frame of the kpc scale jet.  In such a case, the $\gamma$-ray spectra, from the Comptonization of this soft radiation by relativistic electrons, are calculated by integration of equation 3 presented in Banasi\'nski \& Bednarek (2018), see also the formulas in Moderski et al.~(2005) and Aharonian \& Atoyan~(1981). next, these spectra have to be next transformed from the kpc scale jet reference frame to the observer's reference frame according to, 
\begin{eqnarray}
E_\gamma^2 dN/(dE_\gamma dtd\Omega) = D_{\rm obs,z}^4\varepsilon_\gamma^2dN/(d\varepsilon_\gamma dt'd\Omega'),
\label{eq2}
\end{eqnarray}
\noindent
where $dN/(d\varepsilon_\gamma dt'd\Omega'$) is the $\gamma$-ray spectrum in the reference frame of the  kpc scale jet. $D_{\rm obs,z} = [\gamma_{\rm z}(1 - \beta_{\rm z}\cos\theta)]^{-1}$ is the Doppler factor of the kpc scale jet as seen by the external observer. The photon energy in the observer's reference frame is then $E_\gamma = D_{\rm obs,z}\varepsilon_\gamma$.
 
The basic features of the TeV $\gamma$-ray emission, in the case of highly inclined jets in radio galaxy Cen~A,  were discussed in Bednarek (2019). Here we concentrate on the geometrical structure of such TeV $\gamma$-ray emission from the kpc scale jets in the case of more realistic models for the jet deceleration. In fact, some evidence of the extended TeV $\gamma$-ray emission along the kpc scale jet in the nearby radio galaxy, Cen~A, has been  
recently reported by the H.E.S.S. Collaboration (Sanchez et al.~2018). Detailed studies of the morphology of such as $\gamma$-ray emission with the future Cherenkov Telescope Array (CTA) should provide strong constraints on the high energy radiation processes in the kpc scale jets and on the physics of plasma within the jet.

\section{Basic features of the $\gamma$-ray emission}

\begin{figure*}
\vskip 15.5truecm
\includegraphics{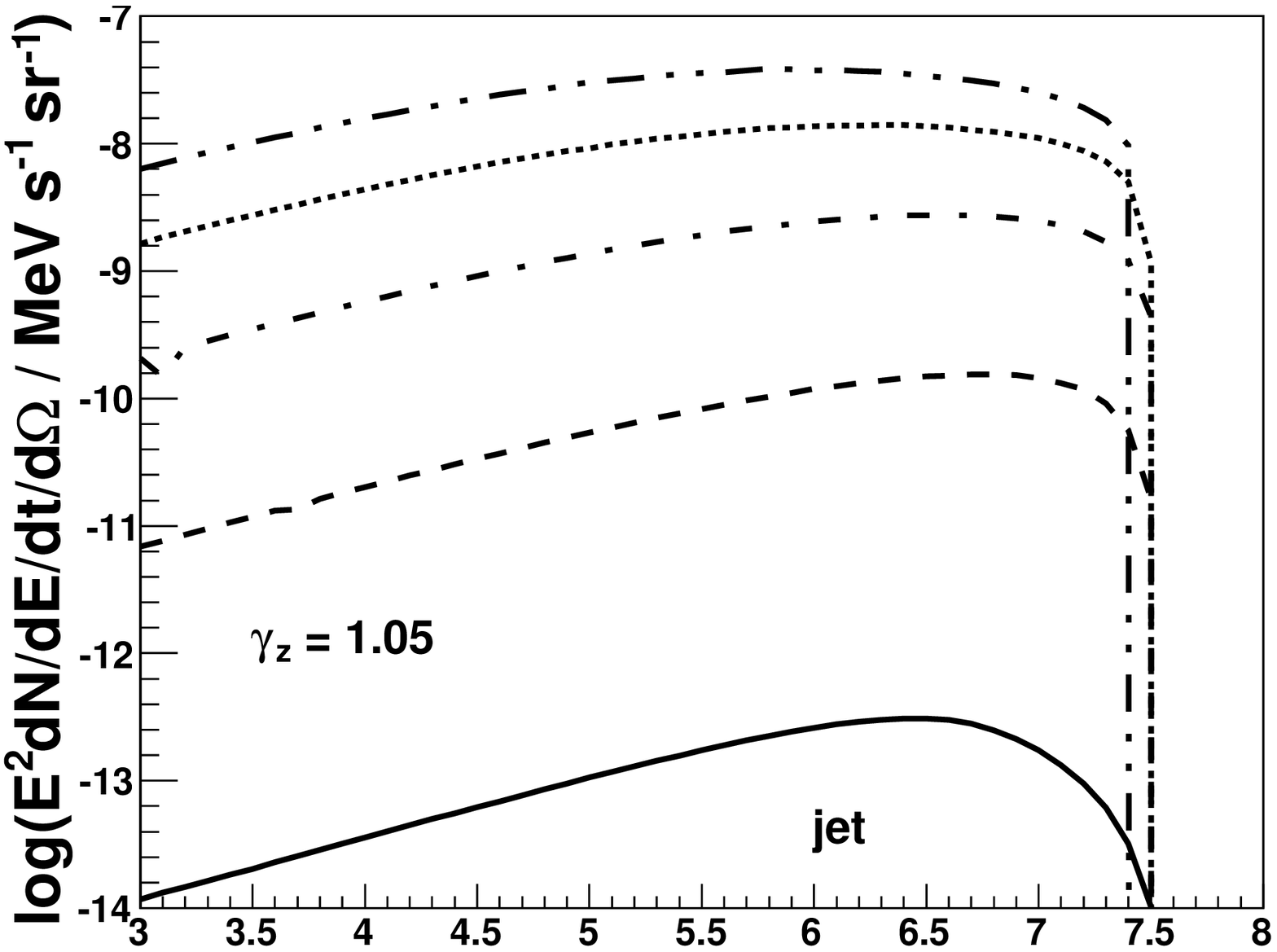}
\includegraphics{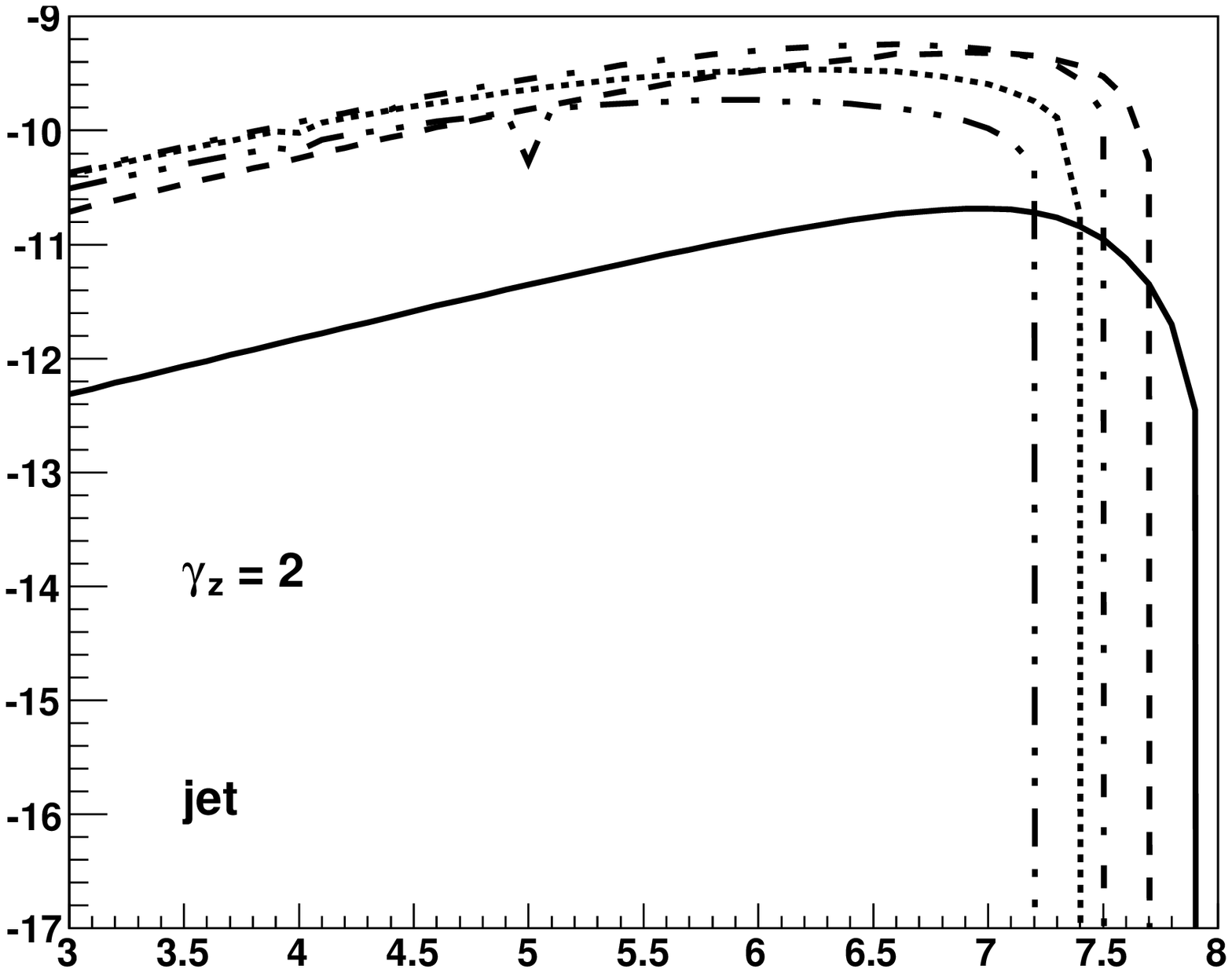}
\includegraphics{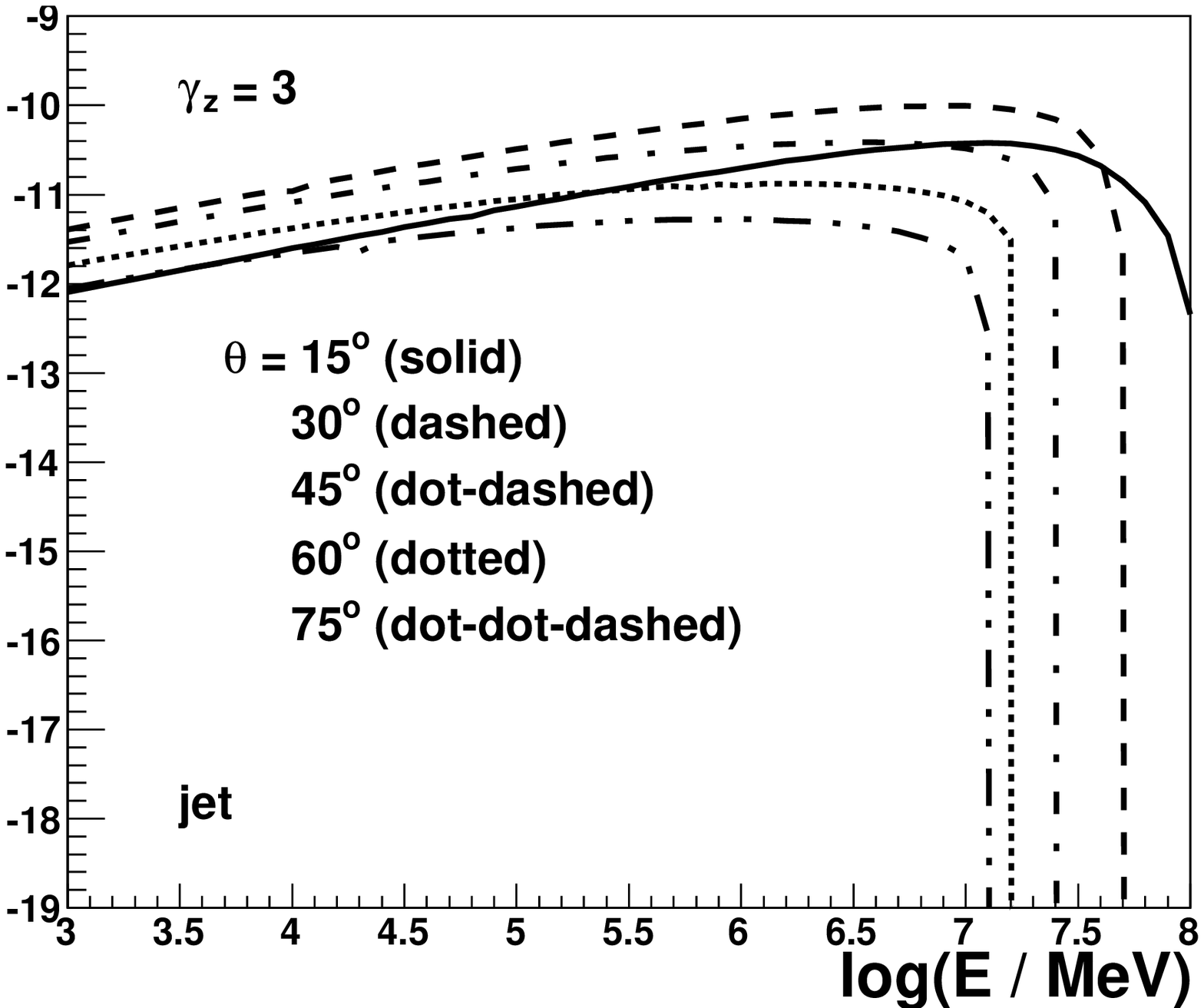}
\includegraphics{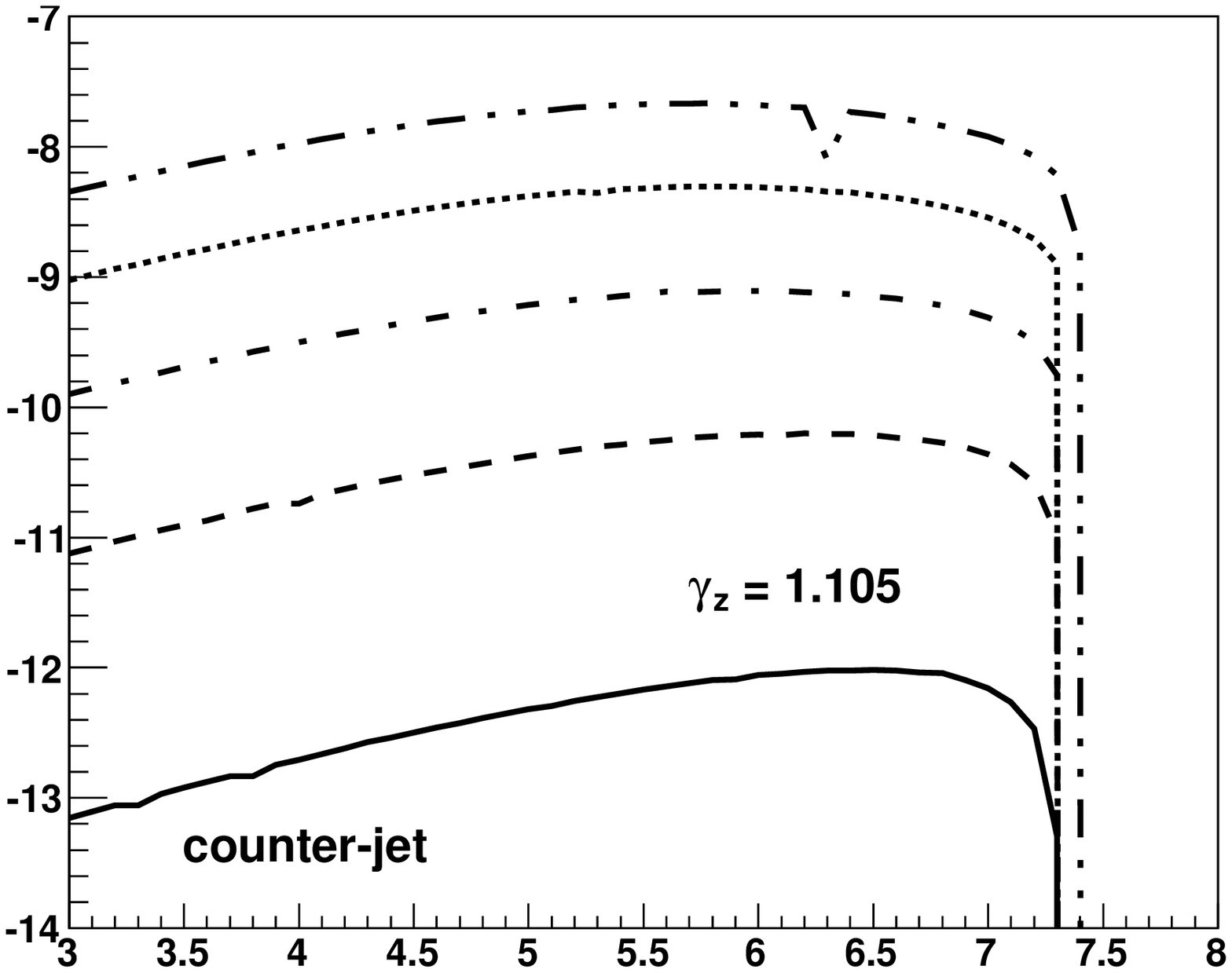}
\includegraphics{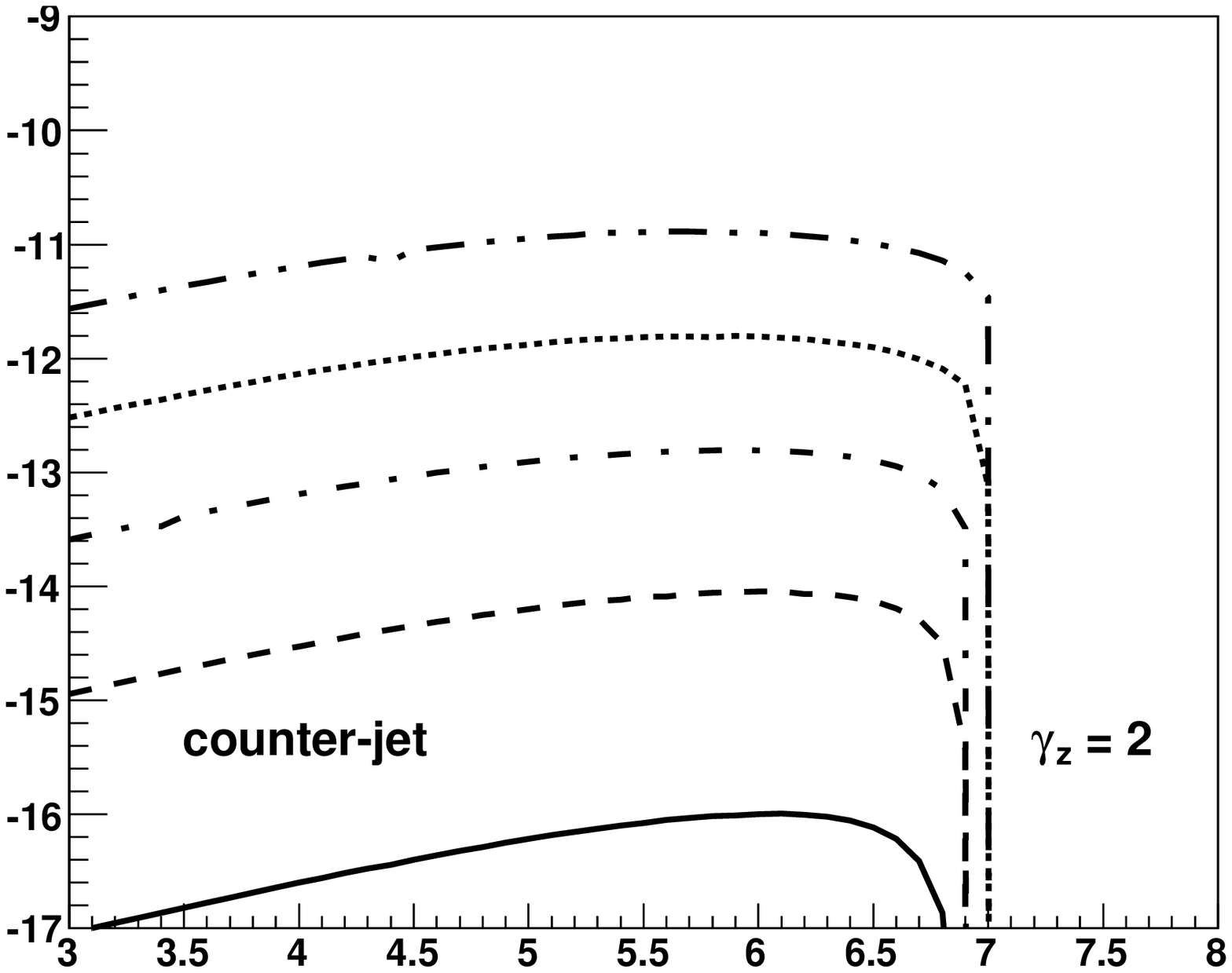}
\includegraphics{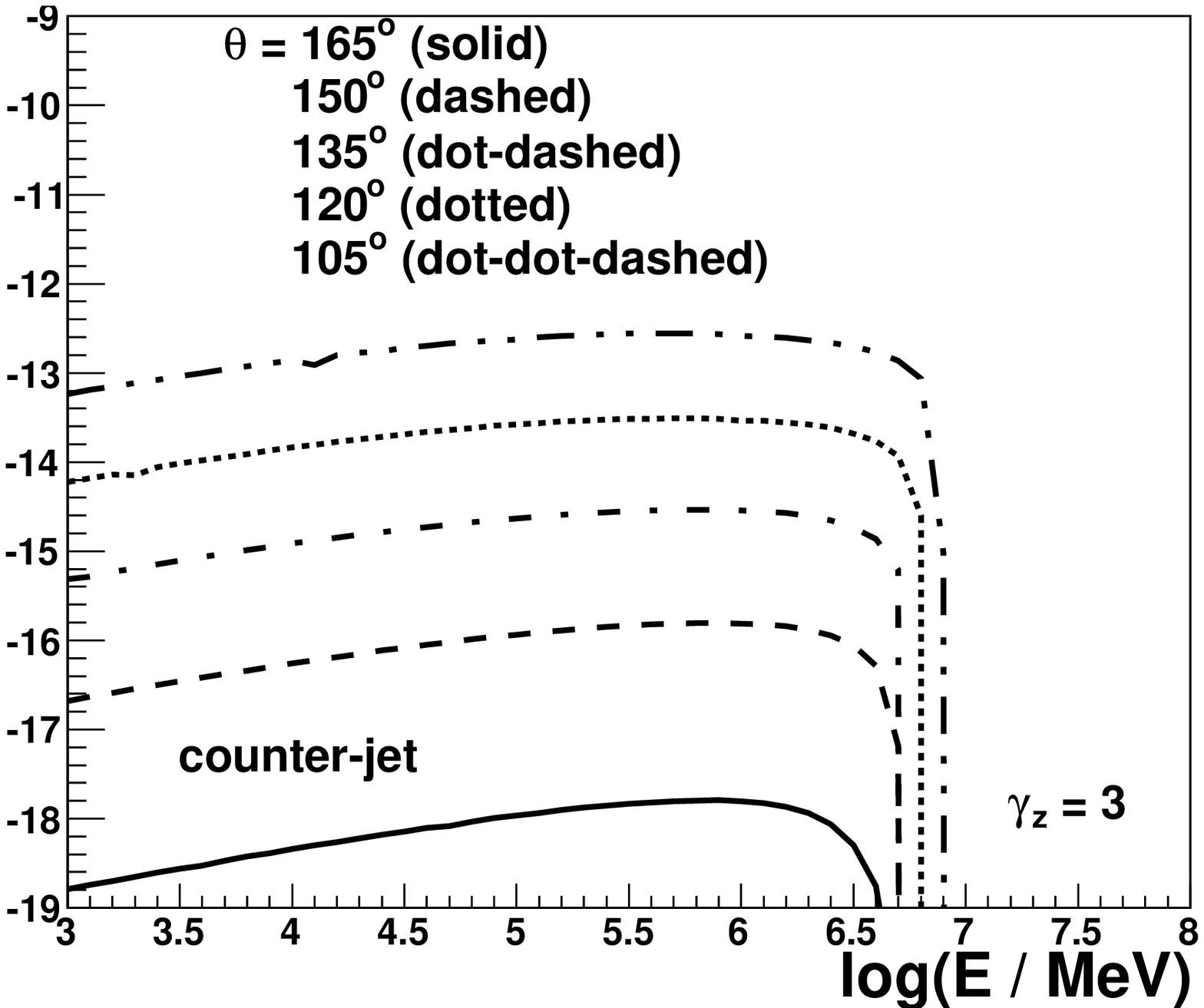}
\caption{Spectral Energy Distribution (SED) of $\gamma$ rays produced in the IC scattering process of relativistic electrons in the kpc scale jet. The electrons scatter soft radiation produced in the inner jet as observed in Cen~A. On the left: The SED is shown as a function of the jet inclination angle $\theta = 15^\circ$ (solid), $30^\circ$ (dashed), $45^\circ$ (dot-dashed), $60^\circ$ (dotted), and $75^\circ$ (dot-dot-dashed). On the right: The SED from the counter-jet is shown for the observation angle equal to $\theta = 165^\circ$ (solid), $150^\circ$ (dashed), $135^\circ$ (dot-dashed), $120^\circ$ (dotted), and $105^\circ$ (dot-dot-dashed). 
The Lorentz factor of the jet is $\gamma_{\rm z} = 1.05$ ($\beta_{\rm z} = 0.3$, top figures), 2 (middle), and 3 (bottom), at the distance from the base of the jet equal to $z = 1$ kpc. The electrons are injected isotropically in the reference frame of the kpc scale jet. They are injected with a power law spectrum with the spectral index 2 between 1 TeV and 30 TeV. The Lorentz factor of the inner jet is $\gamma_{\rm i} = 15$. It is assumed that the observer can only see the soft emission produced in this inner part of the jet which is inclined at smaller angle to the observer's line of sight.} 
\label{fig2}
\end{figure*}

We assume that the kpc scale jet contains isotropically distributed (in the jet frame) relativistic electrons with the equilibrium spectrum well described by a simple power law function, $dN/dE' = A E'^{-\alpha}$, where $A$ is the normalization constant of the spectrum to a single electron, and $\alpha$ is the spectral index. As an example, electrons are assumed to have the spectral index equal to 2 in the energy range between $E'_{\rm min} = 1$ TeV and $E'_{\rm max} = 30$ TeV. In fact, the presence of electrons with multi-TeV energies is required in the kpc scale jet of Cen~A, in order to explain observations of the extended non-thermal X-ray emission detected up to $\sim$7 keV, provided that it has the synchrotron origin (Hardcastle et al.~2003, 2006). In our model, these electrons inverse Compton (IC) up-scatter soft radiation produced in the inner, parsec scale jet of Cen~A.  We have calculated numerically the $\gamma$-ray spectra,
produced in the reference frame of the kpc scale jet, following general formula,  
\begin{eqnarray}
{{dN_\gamma}\over{d\varepsilon_\gamma dt' d\Omega'}} = \int_{E'_{\rm min}}^{E'_{\rm max}} {{dN}\over{dE'}}\times {{dN_\gamma}(E')\over{d\varepsilon_\gamma dt' d\Omega'}} dE',
\label{eq3}
\end{eqnarray}
\noindent
where $dN_\gamma(E')/(d\varepsilon_\gamma dt' d\Omega')$ is the spectrum of $\gamma$-rays produced in a single interaction process by the electron with a constant energy $E'$ (see e.g. Eq.~3 in Banasi\'nski \& Bednarek~2018 and references therein).  
Dependence of the $\gamma$-ray spectrum on the model parameters is investigated in Fig~2. 
Since the inner part of the jet moves with a large Lorentz factor, $\gamma_{\rm i}$, the radiation field produced there has to be strongly Doppler boosted in the reference frame of a relatively slowly moving kpc scale jet. We obtain the $\gamma$-ray spectra in the observer's frame after
applying the transformation given by Eq.~2.  
The $\gamma$-ray spectra from the jet directed towards the observer (and also from the counter-jet), are calculated for a large range of angles $\theta$. We also investigate the dependence of the $\gamma$-ray spectra on the Lorentz factors of the kpc scale jets, starting from the sub-relativistic velocity, $\beta_{\rm z} = 0.3$, and finishing on the mildly relativistic kpc jet moving with the Lorentz factors $\gamma_{\rm z} = 2$ and 3. A few interesting features are evident. For small velocities of the kpc jet, the $\gamma$-ray SED from the jet and the counter-jet are on similar levels. However, the intensity of the  emission clearly increases with the inclination angle of the jet. In the case of small inclination angles, the SED from the counter-jet dominates the one
from the jet itself. The SEDs show different features in the case of mildly relativistic kpc jets. However, the SED does not change drastically with the inclination angle. This is the effect of the interplay between the beaming effect of the radiation along the jet and the specific angular emission pattern of the $\gamma$-rays produced in the reference frame of the kpc scale jet. In fact, the $\gamma$-ray emission pattern has the maximum in the direction of incoming soft radiation, i.e. towards the direction of the inner jet. As expected, the emission from the counter-jet is in this case on a much lower level.

\begin{figure}[t]
\vskip 6.truecm
\includegraphics{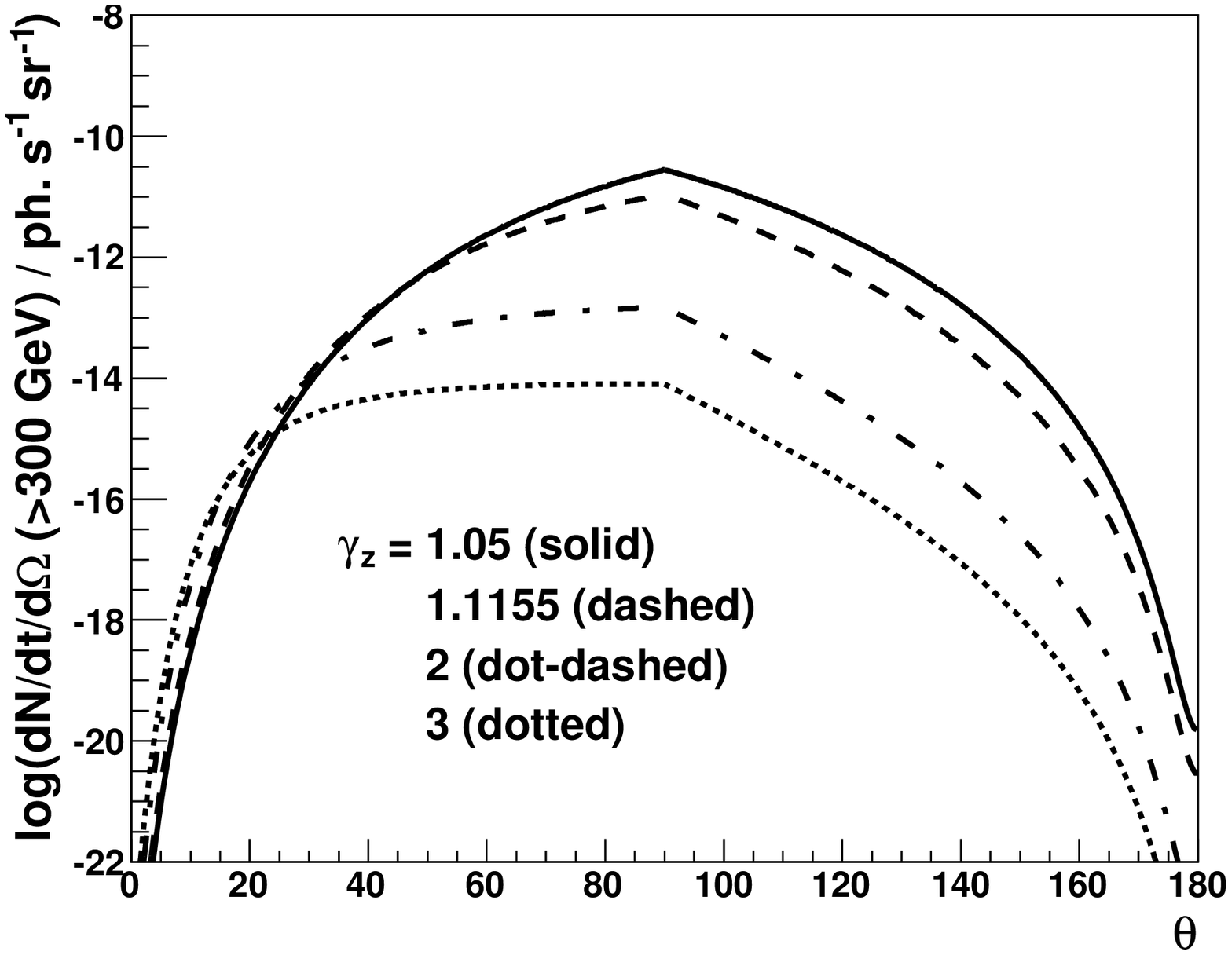}
\caption{The dependence of the $\gamma$-ray flux ($>$300 GeV), produced in the IC process by relativistic electrons for different inclination angles of the jet. The relativistic electrons are isotropically distributed in the reference frame of the kpc scale jet. The electrons have a power law spectrum with the spectral index $2$ between 1 TeV and 30 TeV. They are at the distance of $z = 1$~kpc from the base of the jet. The jet moves with the Lorentz factor $\gamma_{\rm z} = 1.05$ (solid curve, the velocity $\beta_{\rm z} = 0.3c$), 1.1155 (dashed, $\beta_{\rm z} = 0.5c$), 2 (dot-dashed), and 3 (dotted). The inner jet emits soft radiation as observed in Cen~A. This inner part of the jet has a fixed Lorentz factor equal to $\gamma_{\rm i} = 15$.}
\label{fig3}
\end{figure}

For the considered above example parameters of the model, we also calculate the dependence of the $\gamma$-ray flux (above 300 GeV) on the observation angle of the jet (Fig.~3).  The angular distribution of the $\gamma$-ray flux is clearly asymmetric, in respect to the inclination angle $\theta = 90^\circ$
with larger fluxes for the intermediate angles.
This asymmetry effect is stronger for the kpc scale jets moving with larger Lorentz factors due to the relativistic boosting of the produced $\gamma$-ray emission. Moreover, $\gamma$-ray emission close to 
$180^\circ$ is clearly stronger than emission close to  $0^\circ$. This is due to more efficient inverse Compton scattering of radiation into the direction of incoming soft photons. Therefore, emission from the counter-jets, which are viewed at large inclination angles in respect to the observer, is enhanced in respect to the emission from the jets itself. The additional  effect is due to the method of calculation of the soft radiation, from the inner jet/counter-jet, in the location of the kpc scale jet/counter-jet. In order to derive this soft radiation field we assume that the jet and the counter-jet are symmetric. Then, we use the soft emission directly observed in the case of Cen~A. In the case of the jet, the soft radiation is directly observed. But, in the case of the counter-jet ($\theta > 90^\circ$), we assume that the observed soft radiation is viewed at the angle 
($180^\circ - \theta$), i.e. only from the jet itself. From these reasons, the $\gamma$-ray fluxes from the jet and the counter-jet are clearly asymmetric.   

Note also a small kink (at $\theta = 90^\circ$) in the dependencies of the $\gamma$-ray fluxes on the observation angles. It is an artifact of the method which We use in order to derive the soft radiation field in the kpc scale jet. In fact, the observed soft photon spectrum from Cen A is a composition of the emission from the 
inner jet and the inner counter-jet. Therefore, exactly for the angle $90^\circ$, both jets contribute 
with this same level but We only consider soft emission from the jet.  As a result, the applied soft radiation is enhanced by a factor of 2, resulting in the appearance of a small kink at $90^\circ$. This effect decreases fast for the angles smaller and larger than $90^\circ$ but it is difficult to correct due to the complicated dependence of the soft radiation field on the observation angle and on the Lorentz factor of the kpc scale jet. 

\section{The jet with a velocity structure}

\begin{figure}[t]
\vskip 6.truecm
\includegraphics{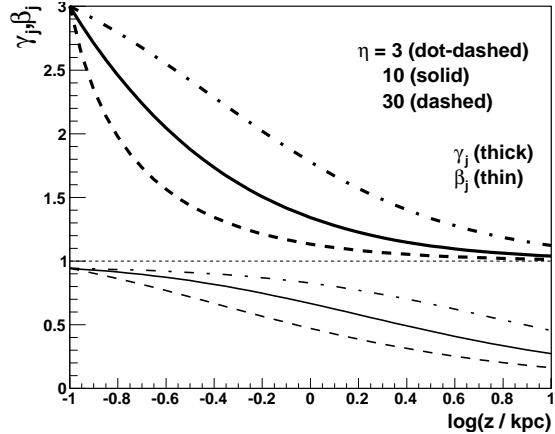}
\caption{The velocity structure of the kpc scale jet ($\beta_{\rm j} = v_{\rm j}/c$) as a function of the distance from the base of the jet, for the deceleration model of the jet defined by the initial Lorentz factor $\gamma_{\rm 0} = 3$ at the distance from the jet base $z_{\rm 0} = 0.1$ kpc, and the deceleration parameter $\eta = 3$ (thin dot-dashed curve), 10 (thin solid), and 30 (thin dashed). The dependence of the Lorentz factors of the jets as a function of distance, $z$, are marked by the corresponding thick curves. The horizontal thin dotted line shows the value of $\gamma_{\rm j}$ and $\beta_{\rm j}$ equal to unity.}
\label{fig4}
\end{figure}

In Fig.~2 we investigated the $\gamma$-ray spectra from the kpc scale jet in the case of a constant 
equilibrium spectrum of electrons in order to evaluate their basic features. 
However, we realize that such idealistic jets do not exist in reality since the jet parameters should change during its propagation. Therefore, we consider more realistic jet models which take into account the effects of in-homogeneity of jets due to 
its interaction with the surrounding medium. 

As we have shown above, the $\gamma$-ray production, considered in terms of our model, strongly depends on the velocity of the jet. Therefore, 
TeV $\gamma$-ray spectra should also strongly depend on this parameter. We introduce a simple model of the jet deceleration on the kpc scale distances, e.g. due to either an entrainment of matter supplied from the medium surrounding the jet or from a large number of stars within the jet. 
In fact, in the case of the jet in Cen~A, the entrainment rate of the matter in the jet, due to winds from the stars, has been estimated on $2.3\times 10^{-3}$~M$_\odot$~yr$^{-1}$ (Wykes et al.~2015). It was commented that such amount of matter is enough to significantly decelerate the jet. However, kpc scale jets in radio galaxies can still move relativistically as indicated by the large jet/counter-jet brightness ratios
in some objects (e.g. Bridle et al.~1994, Wardle \& Aaron 1997). 
At first approximation, we assume that the spectrum of the electrons within the kpc jet is independent on the distance from its base.
On the other hand, the jet velocity, at different distances from the base of the jet, is calculated by assuming that the initial energy of the jet is mainly in the form of a bulk motion of hadronic plasma. We assume that at a certain distance, $z_{\rm 0}$, a mass of 
$M_{\rm 0}$ is ejected into the jet. The initial Lorentz factor of the jet, at $z_{\rm 0}$, can be obtained from $\gamma_{\rm 0} = E_{\rm 0}/M_{\rm 0}c^2$, where $E_{\rm 0}$ is the initial energy injected into the jet, and $\gamma_{\rm 0}$ is its Lorentz factor. Applying the conservation of the momentum of the amount of the mass $M_{\rm 0}$ during its propagation in the jet, the Lorentz factor of the jet is expected to evolve with the distance from the jet base according to $E^2_{\rm 0} - M_{\rm 0}^2c^4 = E^2_{\rm z} - M_{\rm z}^2c^4$, i.e. 
\begin{eqnarray}
\gamma_{\rm z} = [1 + (\gamma_{\rm 0}^2 - 1)M_{\rm 0}/M_{\rm z}]^{1/2}.
\label{eq4}
\end{eqnarray}
\noindent
Due to the entrainment of the background matter into the jet, the initial mass of the jet increases significantly. We assume that in the first approximation the entrainment rate of the matter into the jet is independent on the distance $z$. It is described by 
$M_{\rm z} = M_{\rm 0} + \dot{M}(z-z_0) = M_{\rm 0}[1 + \eta(z-z_0)]$, where $\dot{M} = dM/dz$ is the rate of injection of the background matter into the jet. We mark the ratio of the entrainment rate to the initial mass by $\eta = \dot{M}/M_{\rm 0}$~kpc$^{-1}$. Then, the Lorentz factor of the jet, as a function of the distance $z$ from the jet base, is
\begin{eqnarray}
\gamma_{\rm z} = [1 + (\gamma_{\rm 0}^2 - 1)/(1 + \eta (z-z_0))]^{1/2}.
\label{eq5}
\end{eqnarray}
For typical parameters considered in this model, $\gamma_{\rm o} = 3$, $z_{\rm o} = 0.1$ kpc, the jet with a total power $E_{\rm o} = 10^{44}$~erg~s$^{-1}$, and the entrainment rate of the matter derived by Wykes et al.~(2015, see above), $\eta$ is estimated on $\sim5$~kpc$^{-1}$.

We calculate the profiles of the Lorentz factor of the jet (and its velocity $\beta_{\rm z} = (\gamma_{\rm z}^2 - 1)^{1/2}/\gamma_{\rm z}$) assuming different initial values for the Lorentz factor, $\gamma_{\rm 0}$, and values of the parameter $\eta$.    
The example models for the evolution of the jet Lorentz factors (and its velocities) are shown for specific initial values of $M_{\rm 0}$, $z_{\rm 0}$, $\gamma_{\rm 0}$ and different values of the parameter $\eta$ (see Fig.~4). The jet is assumed to be still relativistic at the distance of 0.1 kpc from its base. 
Note, that recent observations of the jet in another nearby radio galaxy, M 87, show superluminal motion at such distance scales (Bradford et al.~2019).
Depending on the model parameters, we predict either strong or mild deceleration of the jet over the kpc scale distances. 
Confrontation of these predictions with the observations of the distribution of the $\gamma$-ray emission along the jet will allow to put constraints on the jet parameters in the future.
In fact, the prescription for the deceleration of the jet can become quite complicated 
since in this simple model we assume continuous deceleration rate of the jet. For large distances from the base of the jet, the entrainment rate of background matter into the jet likely drops. Then, the jet stabilizes at sub-relativistic velocities.

\begin{figure*}
\vskip 12.5truecm
\includegraphics{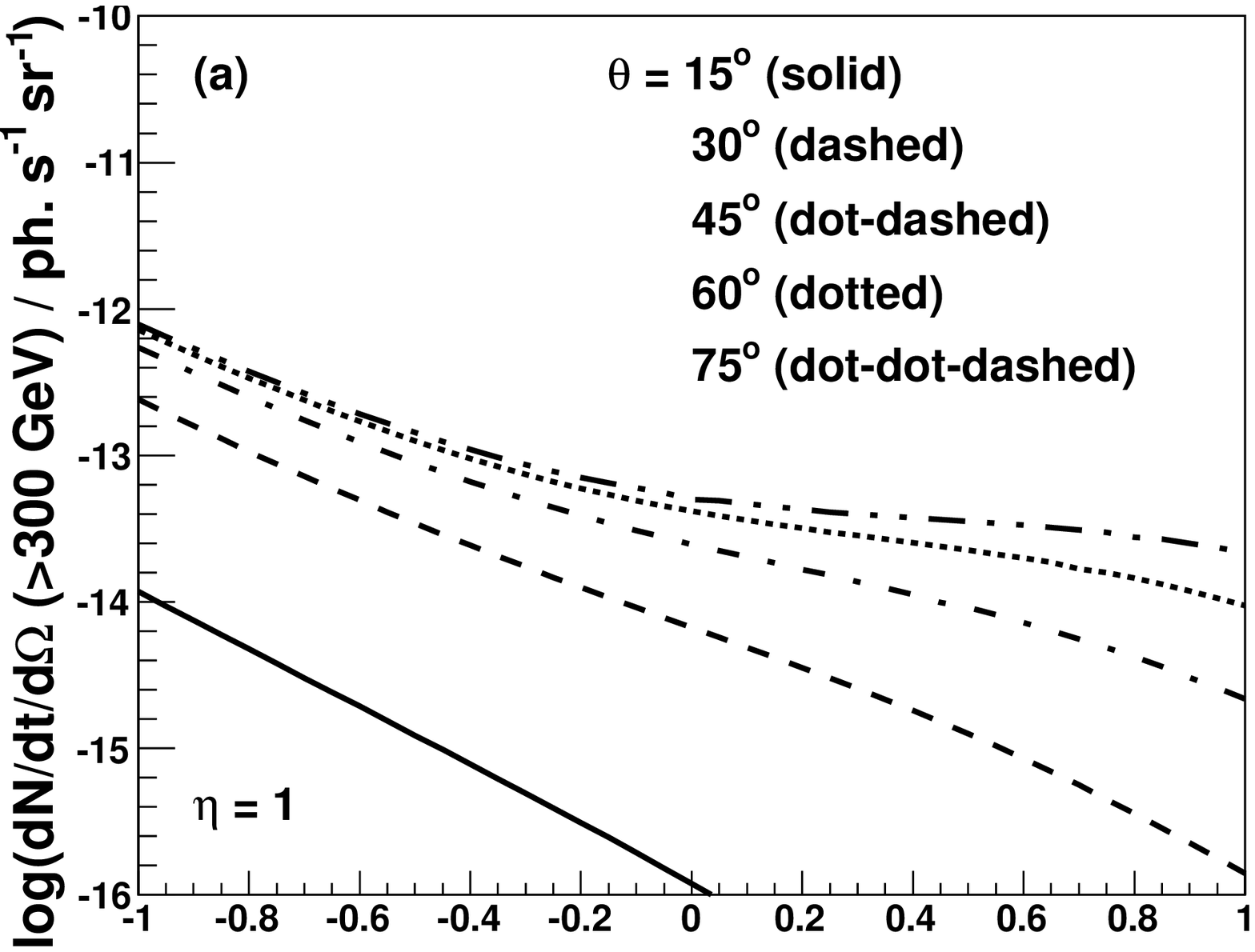}
\includegraphics{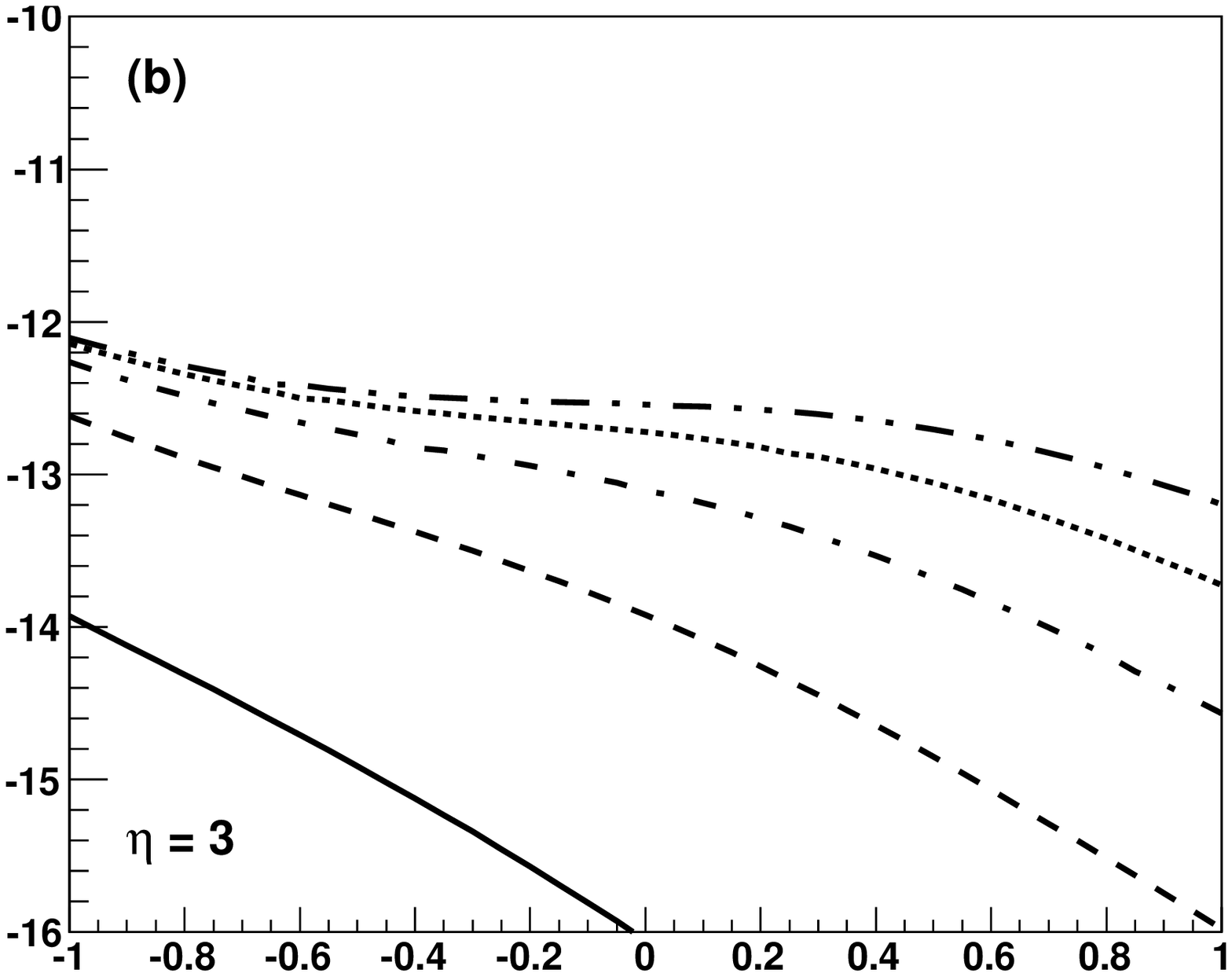}
\includegraphics{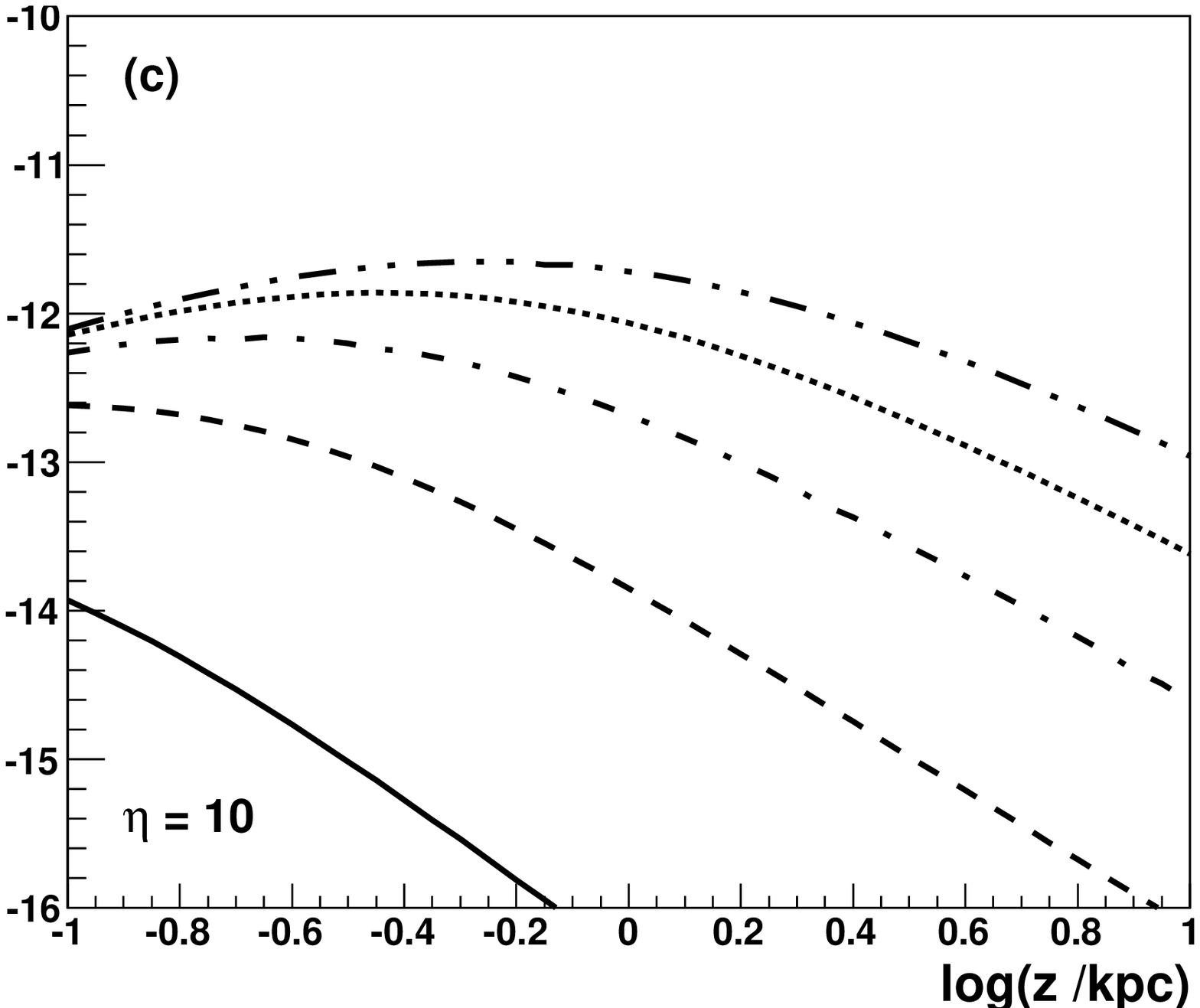}
\includegraphics{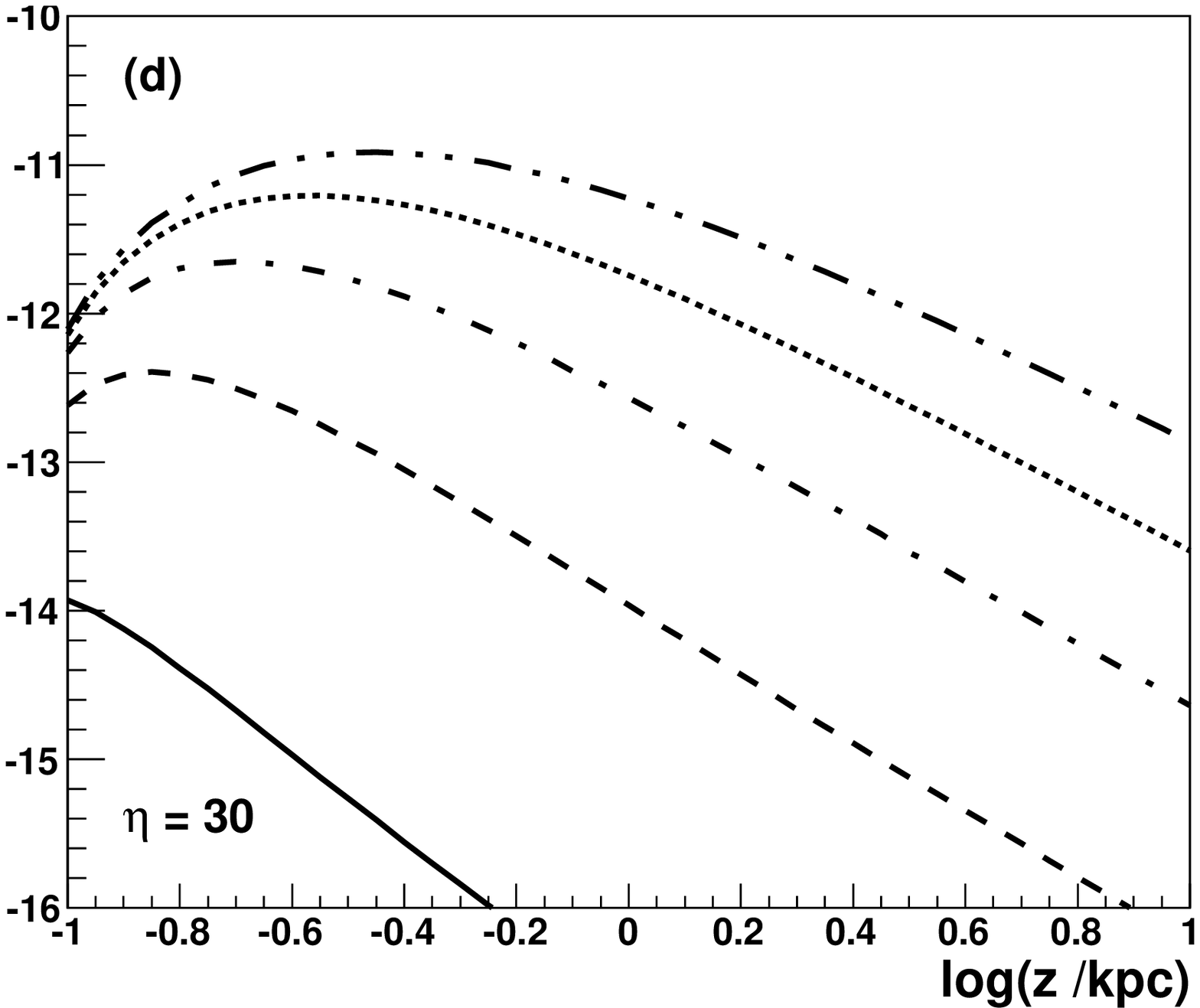}
\caption{Distribution of the $\gamma$-ray photon flux ($>$300 GeV) along the kpc scale jet as a function of the real distance along the jet. Different deceleration models of the jet, defined by the initial jet Lorentz factor equal to $\gamma_{\rm 0} = 3$ (at $z_{\rm 0} = 0.1$ kpc) and the parameter describing deceleration of the jet $\eta = 1$ (a), 3 (b), 10 (c), and 30 (d), are considered. The jet is viewed at the angle $\theta = 15^\circ$ (solid), $30^\circ$ (dashed), $45^\circ$ (dot-dashed), $60^\circ$ (dotted), and $75^\circ$ (dot-dot-dashed). The injection spectra of relativistic electrons, and the soft radiation field from the inner jet, are discussed in Sect.~3. It is assumed that the electron injection spectrum is independent on the distance from the base of the jet.} 
\label{fig5}
\end{figure*}

Applying the above defined simple model for the velocity structure of the jet, we calculate the $\gamma$-ray fluxes (above 300 GeV) produced by relativistic electrons which IC up-scatter inner jet soft radiation. They are shown for different jet models as a function of the distance (measured along the jet) and for a few selected values of the observation angles of the jet (see Fig.~5). Specific flux profiles change with the value of the parameter $\eta$ describing the deceleration of the jet, $\eta = 1$ (figure a), 3 (b), 10 (c), and 30 (d). For large viewing angles, and a relatively small $\eta$, 
the $\gamma$-ray flux extends with similar level up to several kpc from the jet base. For fast decelerated jets (figures c and d), the TeV $\gamma$-ray flux reaches a broad maximum at sub-kpc distances. This effect is due to  the
combination of efficient deceleration of the jet and weakening of the soft photon field from the inner jet with the distance from its base. The $\gamma$-ray flux drops fast with the distance from the jet base for small observation angles. This effect is mainly due to the decay of the soft radiation from the inner jet with the distance from the jet base (as $\propto d^{-2}$). We conclude that the model predicts significant TeV $\gamma$-ray fluxes at a kpc scale distances from the base of the jet for relatively large observation angles ($>$40$^\circ$).

\section{Gamma-rays from electrons evolving within the jet}

\begin{figure*}
\vskip 12.truecm
\includegraphics{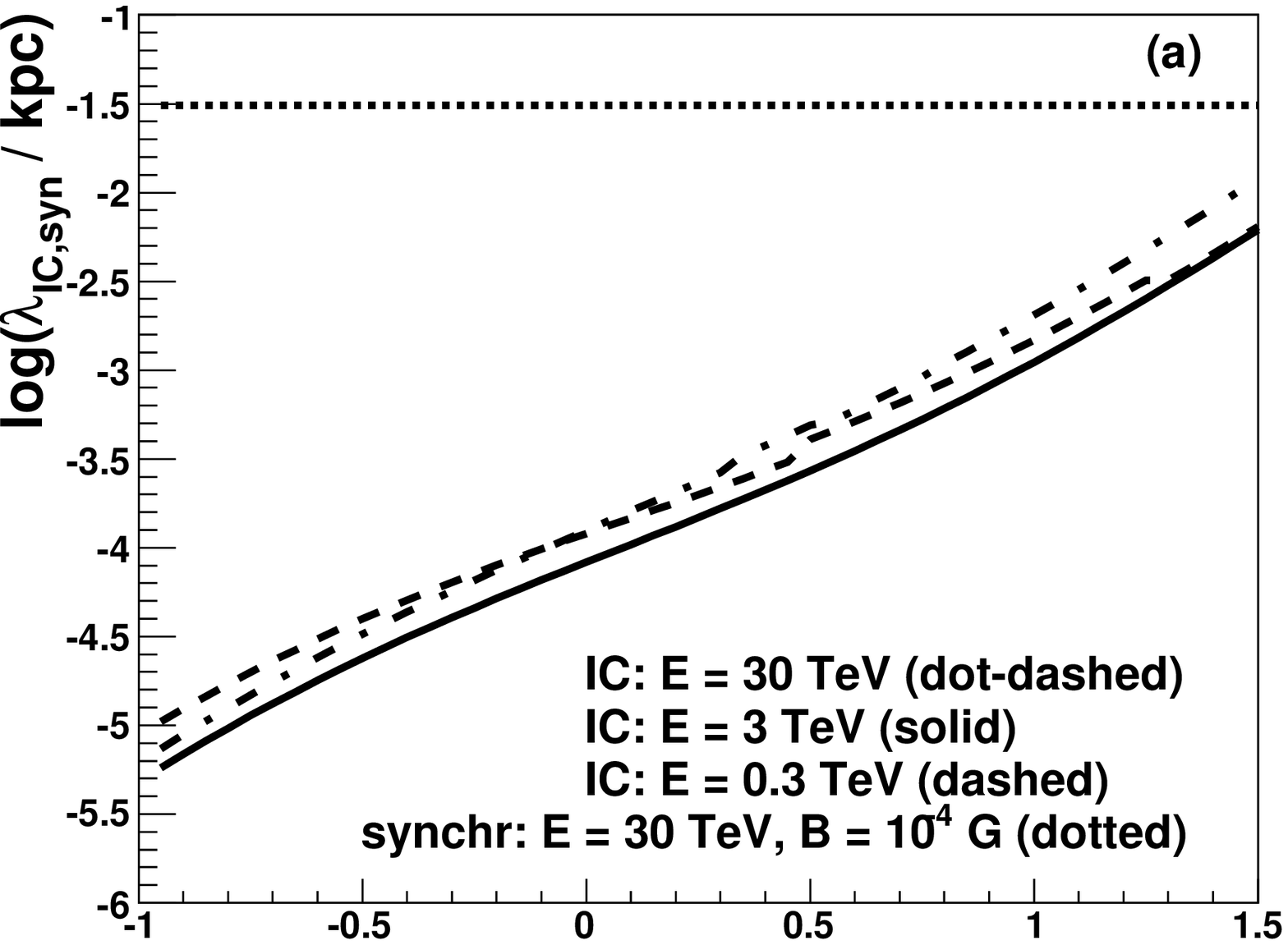}
\includegraphics{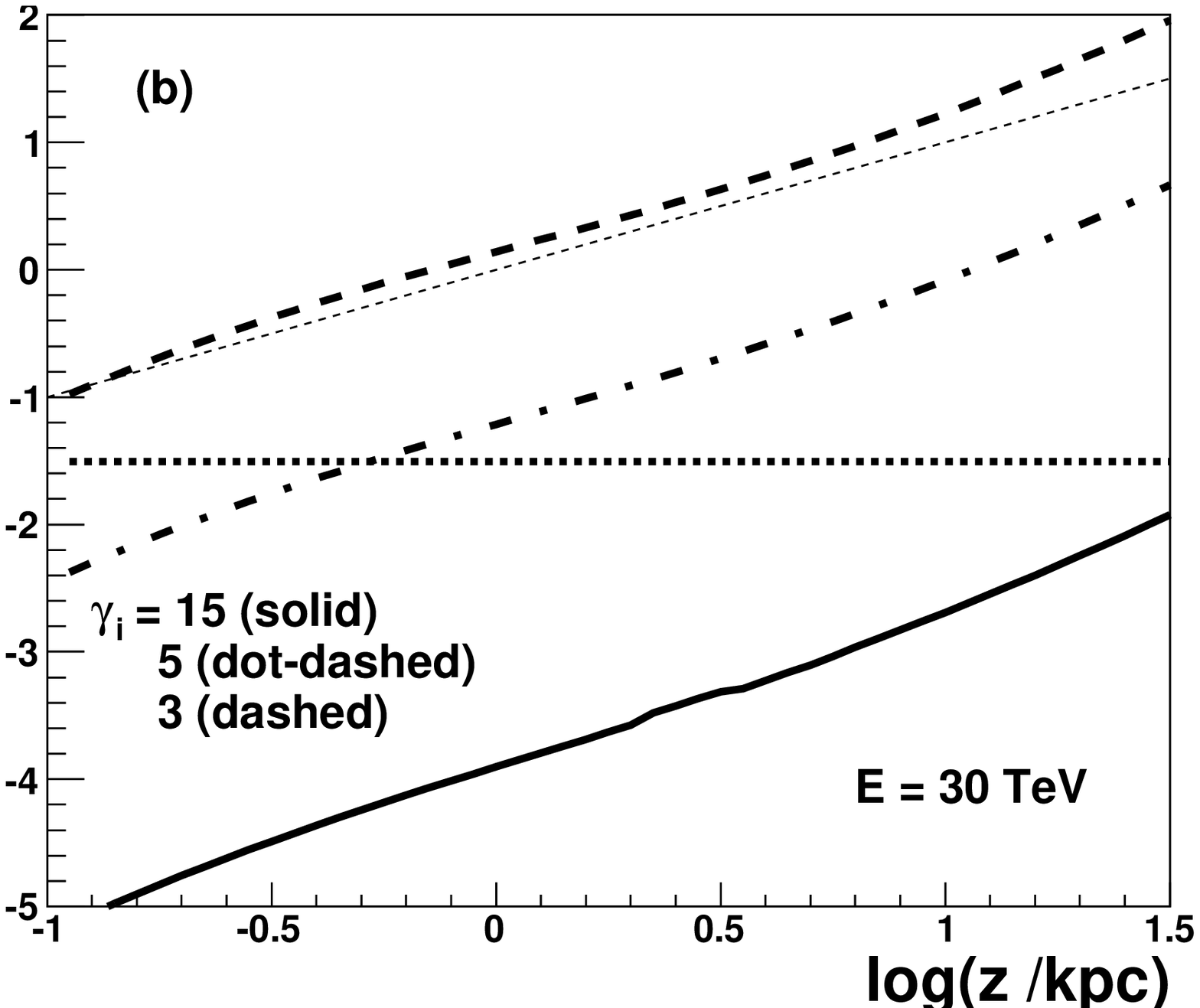}
\includegraphics{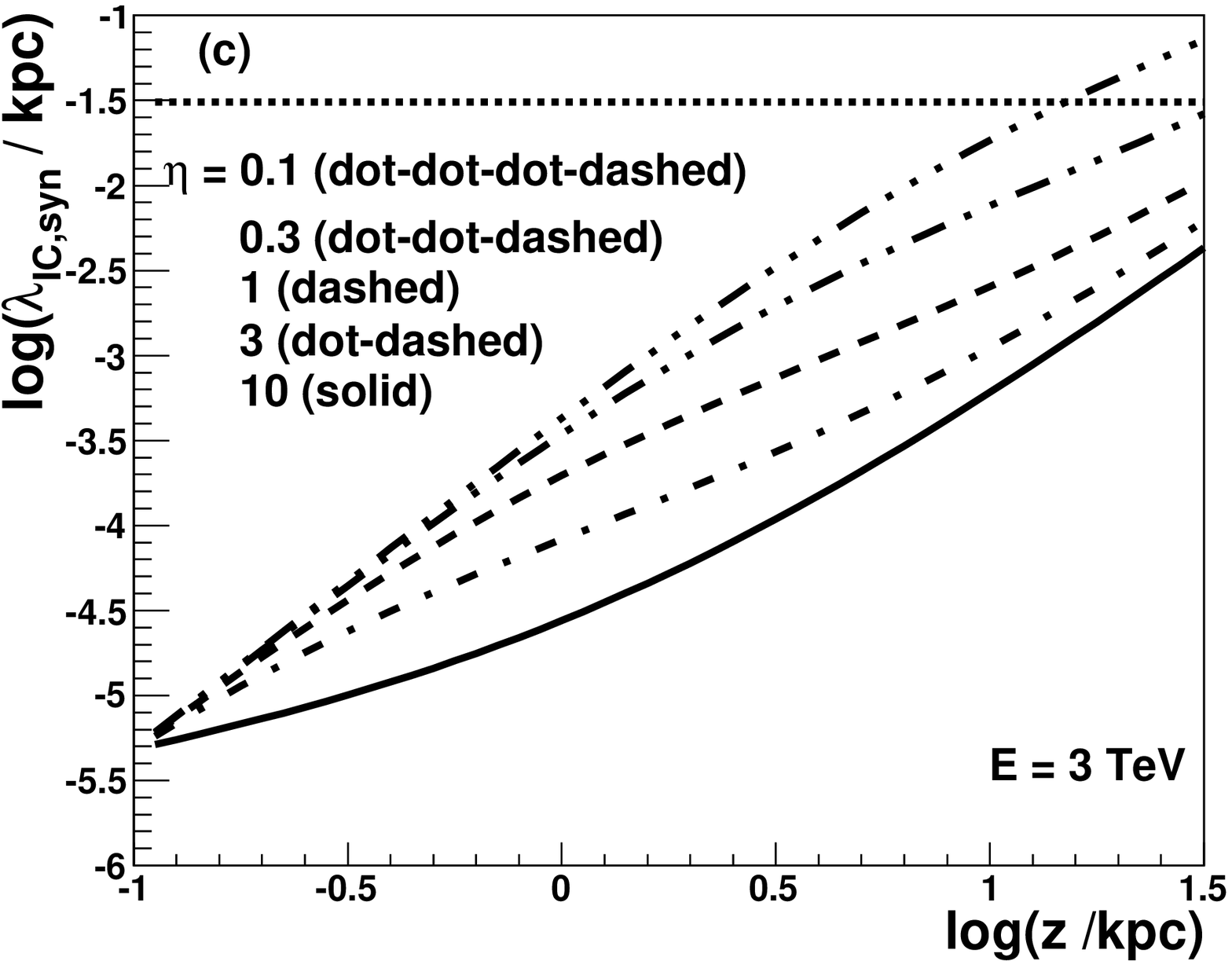}
\includegraphics{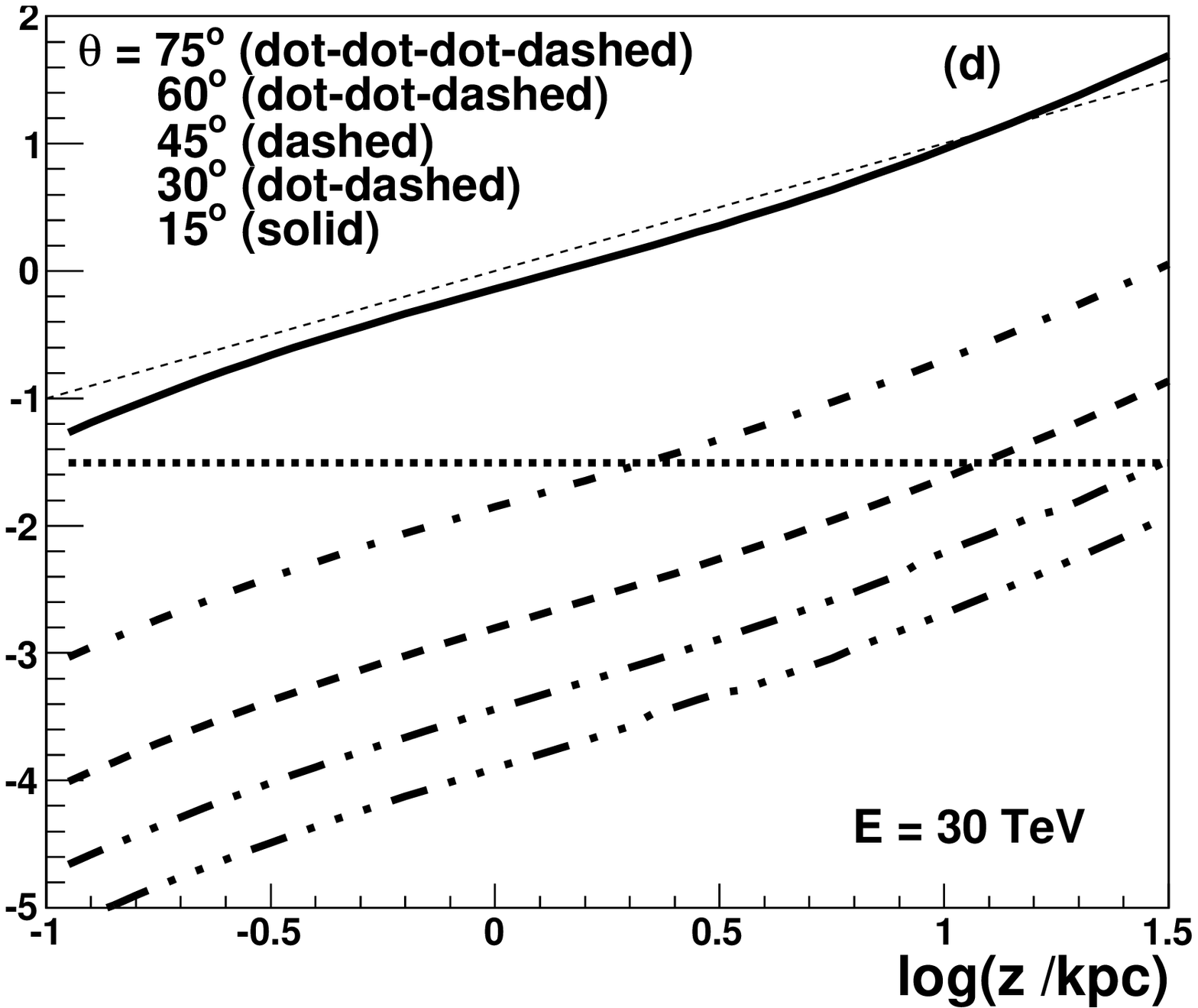}
\caption{The mean free paths for relativistic electrons due to their energy losses on the inverse Compton process, occurring within the large scale jet as a function of the distance from the base of the jet. The radiation field is produced in the inner jet as described in the text. The results are shown for selected energies of the electrons (figure a), the Lorentz factors of the inner jet (b), the parameter $\eta$ (which determine the evolution of the jet) and the inclination angles of the jet $\theta$. In (a): the electron energy is $E = 30$ TeV (dot-dashed curve), 3 TeV (solid), and 0.3 TeV (dashed) 
for the observation angle $\theta = 75^\circ$. The mean free path on the synchrotron process, for a magnetic field of $B = 10^{-4}$~G, is marked by the horizontal thick dotted curve for $E = 30$ TeV. In all calculations we use the values $\gamma_{\rm i} = 15$, $\gamma_{\rm 0} = 3$, $\eta = 3$ (unless specified differently). In (b): the inner jet Lorentz factor is $\gamma_{\rm i} = 15$ (solid), 5 (dot-dashed), and 3 (dashed), and the electron energy is 30 TeV. In (c): the parameter $\eta = 10$ (solid), 3 (dot-dashed), 1 (dashed), 0.3 (dot-dot-dashed), 0.1 (dot-dot-dot-dashed), and the electron energy is 3 TeV. In (d): the inclination angle $\theta$ is $15^\circ$ (solid), $30^\circ$ (dot-dashed), $45^\circ$ (dashed), $60^\circ$ (dot-dot-dashed), and $75^\circ$ (dot-dot-dot-dashed), the electron energy is 30 TeV. The case $\lambda_{\rm IC} = z$ is shown in (b) and (d) by the thin dotted line.} 
\label{fig6}
\end{figure*}

In this chapter, we consider more complete model in which the electrons are injected into the jet with a constant rate and a power law spectrum. These electrons reach local equilibrium due to the energy losses. We assume that relativistic electrons suffer energy losses mainly on the synchrotron process, in the magnetic field of the large scale jet, and on the Inverse Compton scattering of soft radiation, produced in the inner, fast moving, parsec scale jet. 

\subsection{Cooling of the electrons}

In order to conclude whether the electrons cool locally in the jet or they are at the same time effectively advected along the jet, we calculate the mean free paths, $\lambda = -cE/\dot{E}$, for relativistic electrons on the IC and synchrotron processes. The average IC energy losses, $\dot{E}_{\rm IC}$, are calculated in the general case by integrating the spectrum of produced $\gamma$-rays (see Eq. 2.48 in Blumenthal \& Gould~1970). The synchrotron energy losses are $\dot{E}_{\rm syn} = dE/dt' = -(4/3)c\sigma_{\rm T}\rho_{\rm B}\gamma_{\rm e}^2$, where
$\sigma_{\rm T}$ is the Thomson cross section, $\rho_{\rm B}$ is the energy density of the magnetic field in the jet at the distance $z$ from its base, and $\gamma_{\rm e}$ is the Lorentz factor of the electrons. 
We investigate the dependence of $\lambda_{\rm IC} = -cE/\dot{E}_{\rm IC}$ on the parameters of the model, such as the electron energy, $E$, the Lorentz factor of the inner jet, $\gamma_{\rm i}$, the parameter $\eta$ describing the longitudinal evolution of the jet, and the observation angle of the jet, $\theta$, which determine the strength of the radiation field produced in the inner jet (see Fig.~6). In those calculations, the type of the soft radiation field as observed from Cen~A is applied, assuming that the inner jet moves with the Lorentz factor $\gamma_{\rm i}$ (see Sect.~2). For such a radiation field, $\lambda_{\rm IC}$ increases proportionally with the distance measured along the jet (Fig.~6). Note also, that 
$\lambda_{\rm IC}$ depends only weakly on the energy of the injected electrons which is due to the shape of the applied soft radiation field (Fig.~6a). The electrons scatter the soft radiation from the inner jet in both the Thomson and in the Klein-Nishina regime. The dependence of $\lambda_{\rm IC}$ on the electron's energy in those two regimes is inverted.
Therefore, it is not surprising that for specific parameters and different distances along the jet, 
$\lambda_{\rm IC}$ shows complicated dependence on the electron energy. For example, at the base of the jet, electrons with energy equal to 30 TeV have larger $\lambda_{\rm IC}$ than electrons with energy equal to 3 TeV. But, this dependence inverts to just the opposite at large distances from the jet base when the jet movement is already sub-relativistic. 

On the other hand, $\lambda_{\rm IC}$ depends strongly on the Lorentz factor of the inner jet. This Lorentz factor   determines the boosting effect of soft radiation along the jet axis (Fig.~6b). The velocity structure of the kpc scale jet has also an important effect on the energy losses of the electrons. $\lambda_{\rm IC}$ becomes clearly shorter for  jets effectively decelerated (i.e. for large values of the parameter $\eta$). 
We also observe interesting dependence of 
$\lambda_{\rm IC}$ on the observation angle of the jet (see Fig. 6d). Above mentioned effects are caused by the strong dependence of the boosting effect of the soft radiation from the inner jet, as observed at the distance of the kpc scale jet, on the velocity of the kpc scale jet and on its observation angle. For reasonable range of the model parameters, we have obtained that $\lambda_{\rm IC}$ is usually clearly shorter than the local distance scale of the jet (see inclined thin dotted line in Figs. 6b and 6d). 
We study the condition for the local versus non-local cooling of electrons in the jet in a more detail in Fig.~7. In this figure we plot the range of values 
($\gamma_{\rm i}$, $\theta$) for which $\lambda_{\rm IC}$ becomes comparable to the distance from the base of the jet, i.e.  $\lambda_{\rm IC} = z$.  
It is clear that even for mildly relativistic inner jets, $\gamma_{\rm i} > 10$, the relativistic electrons usually cool locally in the jet. Then, the effects, related to the advection of the electrons along the jet, can be safely neglected for a reasonable range of model parameters. We also compare $\lambda_{\rm IC}$ with the energy losses  of the electrons on the synchrotron process for the magnetic field strength within the kpc scale jet equal to $B = 10^{-4}$ G (see horizontal thick dotted lines in Figs.~6). The synchrotron energy losses of the electrons are usually lower than their energy losses on the IC process. Therefore, more energy is transferred from relativistic electrons to the high energy $\gamma$ rays (IC process) than to the energy range from radio up to X-rays (synchrotron radiation). The synchrotron radiation dominated jets can be observed in the case of a relatively slow inner jets (Lorentz factor of the inner jet $\gamma_{\rm i}\le 10$) and in the case of a small inclination angles of the jets towards the observer ($\theta\le 45^\circ$).  For such parameters, the soft radiation from the inner jet is relatively weak in the reference frame of the kpc scale jet.         

\begin{figure}
\vskip 9.5truecm
\includegraphics{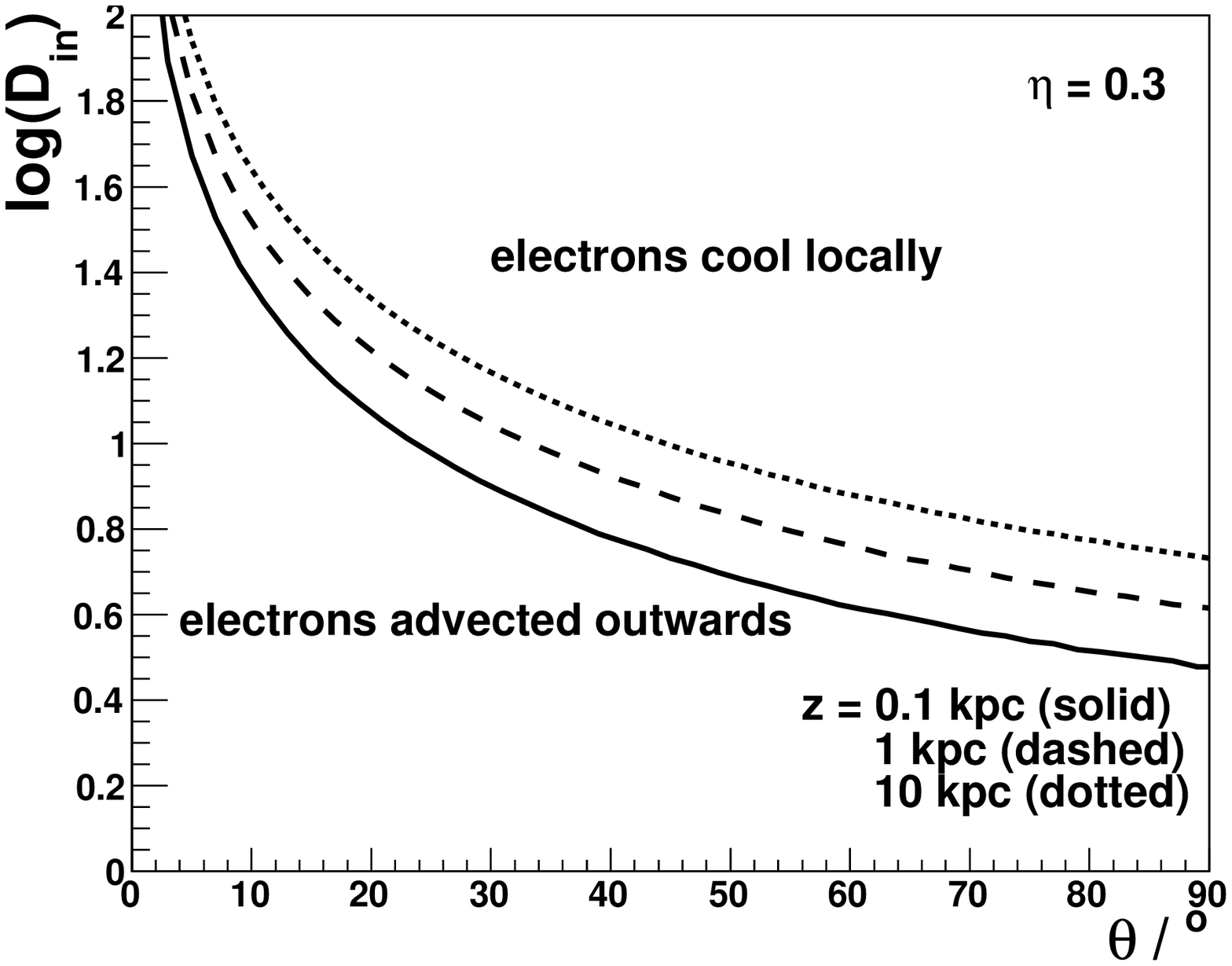}
\includegraphics{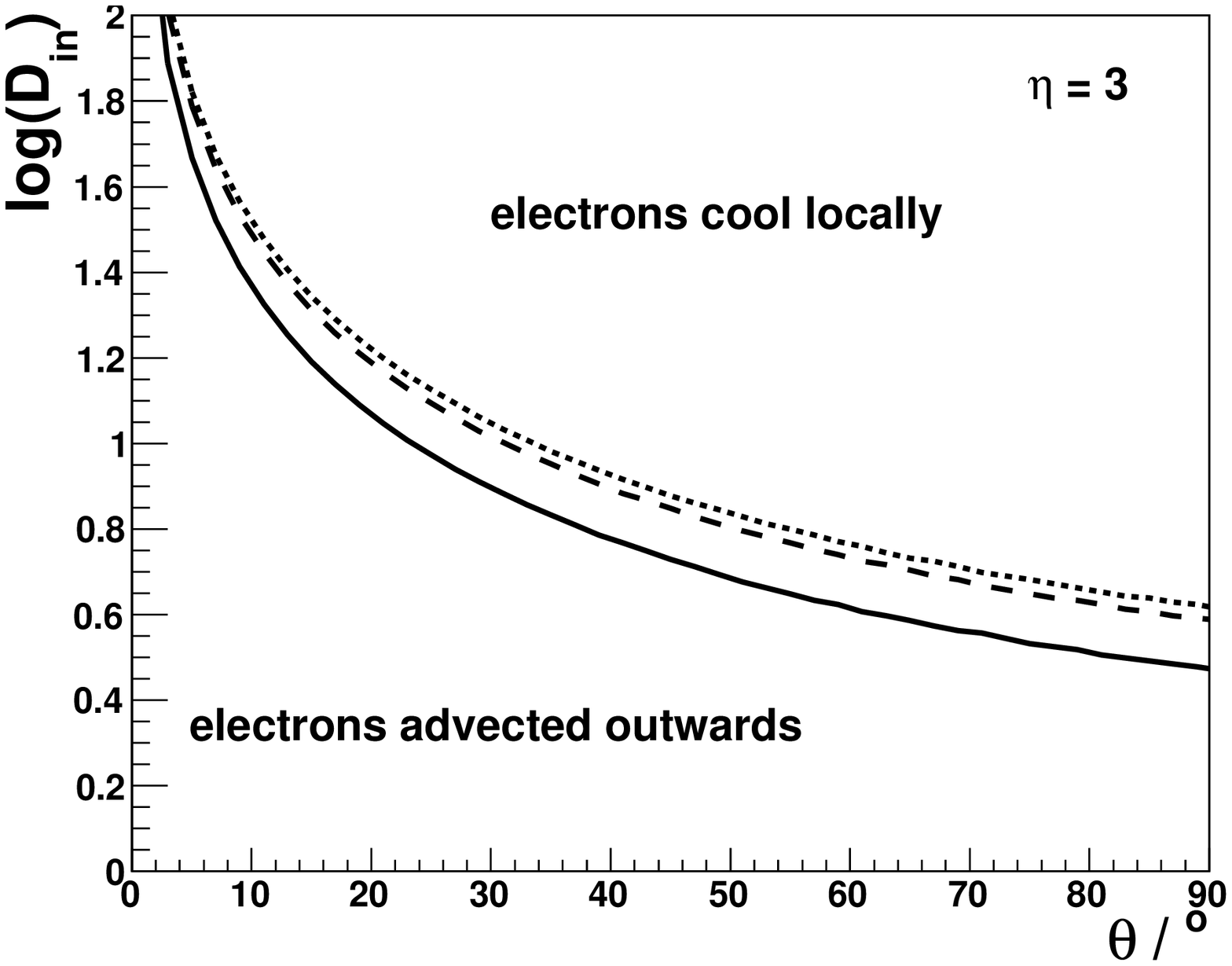}
\caption{The parameter space is shown, log of the Doppler factor of the inner jet versus the observation angle of the jet, for which relativistic electrons either cool locally in the jet or they are advected outward from the jet. It is shown for the parameters $\eta = 0.3$ (top) and 3 (bottom). The specific curves correspond to the following condition:
the local mean free path for relativistic electrons on the IC process is equal to the distance from the base of the jet, i.e. $\lambda_{\rm IC} = z$. These example conditions are shown for the specific locations within the jet defined by the distance 0.1 kpc (solid curve), 1 kpc (dashed), and 10 kpc (dotted) measured from the jet base. The Lorentz factor of the jet is assumed to be equal to $\gamma_{\rm 0} = 3$ at $z_{\rm 0} =  0.1$ kpc and electrons have energy equal to $E = 3$ TeV.}
\label{fig7}
\end{figure}

In the next section, we calculate the $\gamma$-ray spectra produced in the kpc scale jet, assuming 
that the relativistic electrons cool locally in the jet.

\subsection{Distribution of gamma-ray emission along the jet}

We assume that relativistic electrons are injected along the jet (in its reference frame) with a constant rate. Their energy distribution is well described by the power law function,
\begin{eqnarray} 
dQ/(dE'dt'\Delta z) = A_{\rm z}A_{\rm E}E'^{-\alpha}, 
\label{eq5b}
\end{eqnarray}
\noindent
in the energy range from $E'_{\rm min}$ to $E'_{\rm max}$, where A$_{\rm E}$ is the normalization constant obtained for the power in relativistic electrons equal to 1 MeV s$^{-1}$. $A_{\rm z}$ is the normalization of the injection of electrons as a function of distance along the jet equal to 1 kpc$^{-1}$, and $\alpha$ is the spectral index. Since the electrons lose energy dominantly on the IC and synchrotron processes, their equilibrium spectrum at a specific distance, z, from the base of the jet, can be obtained from,
\begin{eqnarray}
{{dN}\over{dE \Delta z}} =\dot{E}^{-1}\int_{\rm E}^{E'_{\rm max}} {{dQ}\over{dE' dt' \Delta z}}dE'.
\label{eq6}
\end{eqnarray}
\noindent
where the energy loss rate of electrons, $\dot{E} = \dot{E}_{\rm IC} + \dot{E}_{\rm syn}$, is defined in Sect.~5.1.
The spectra of $\gamma$ rays, produced within a layer with the thickness, $\Delta z$, located at the distance, z, from the jet base, can be calculated by integrating the following equation, 
\begin{eqnarray}
{{dN_\gamma}\over{d\varepsilon_\gamma dt' d\Omega' \Delta z}} = \int_{E_{\rm min}}^{E'_{\rm max}} {{dN}\over{dE\Delta z}}\times {{dN_\gamma}(E')\over{d\varepsilon_\gamma dt' d\Omega'}} dE,
\label{eq7}
\end{eqnarray}
\noindent
where $dN_\gamma(E')/(d\varepsilon_\gamma dt'd\Omega')$ is the spectrum of $\gamma$-rays produced by electrons with energy $E'$
(e.g. Eq. 3 in Banasi\'nski \& Bednarek~2018).
This $\gamma$-ray spectrum is next transformed to the observer's reference frame by using Eq.~2.
We also calculate the $\gamma$-ray flux, as a function of distance from the base of the jet, z, by integrating the $\gamma$-ray spectrum in the observer's reference frame above the minimum energy $E_\gamma^{\rm min}$.  

\begin{figure*}
\vskip 12.5truecm
\includegraphics{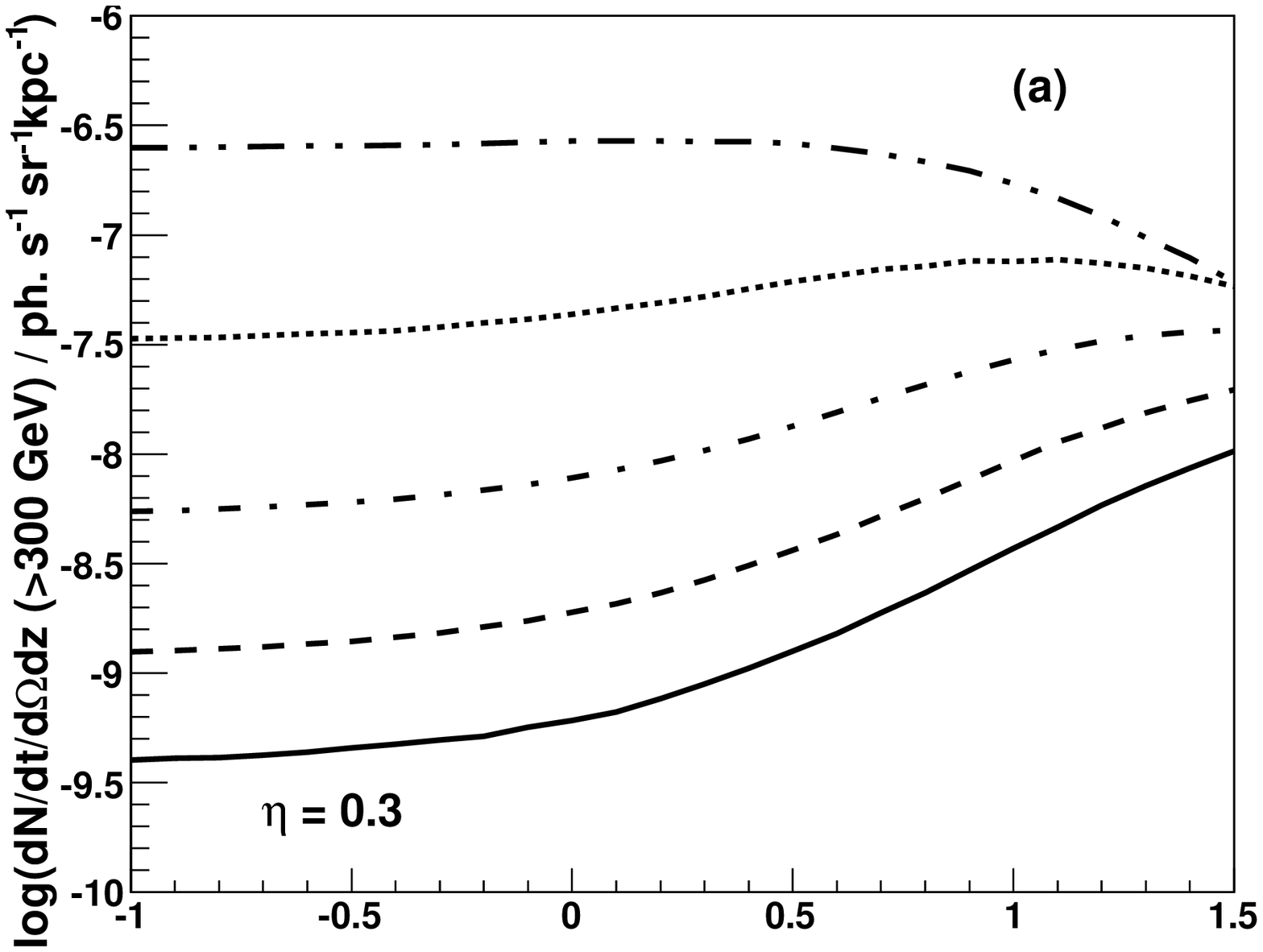}
\includegraphics{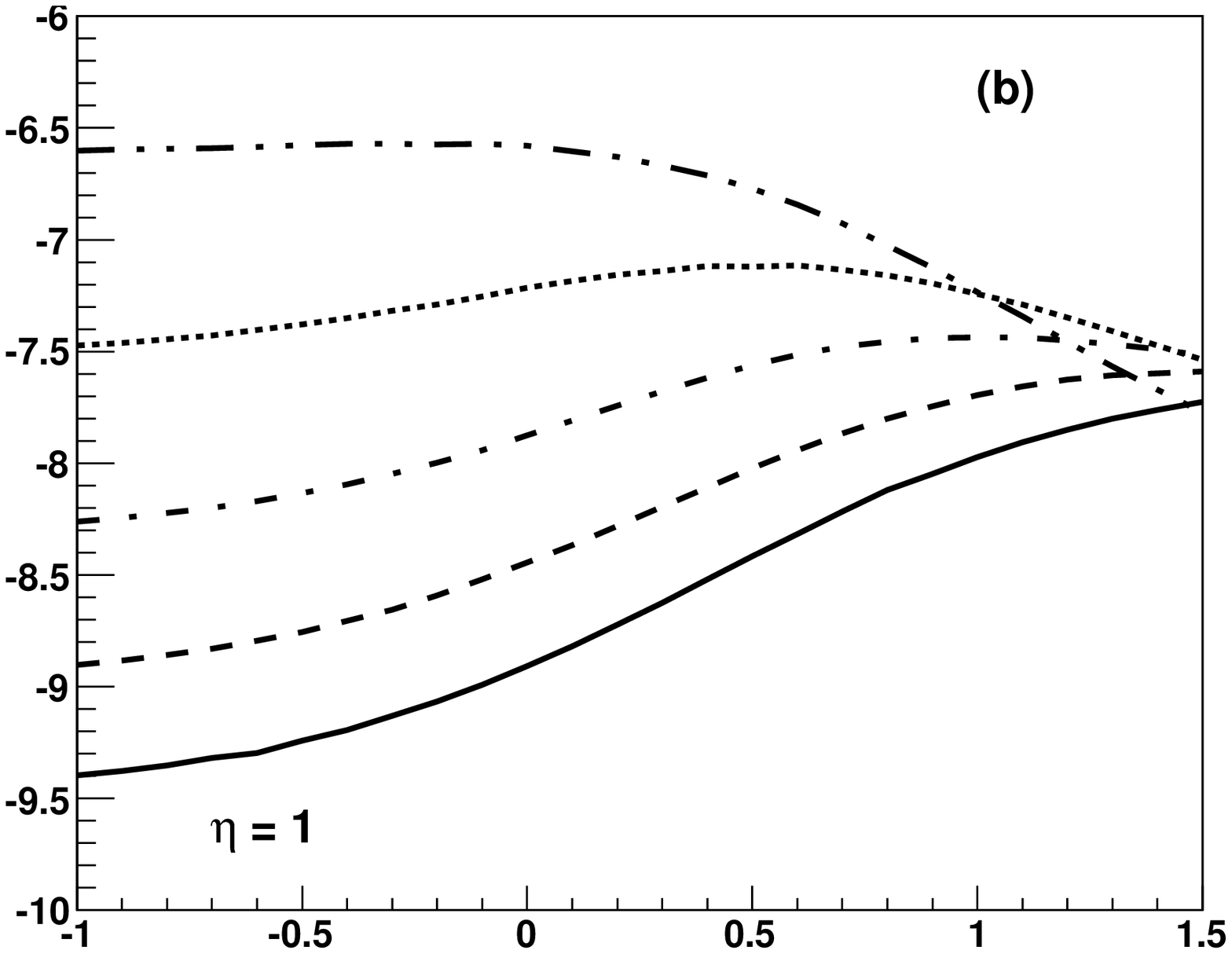}
\includegraphics{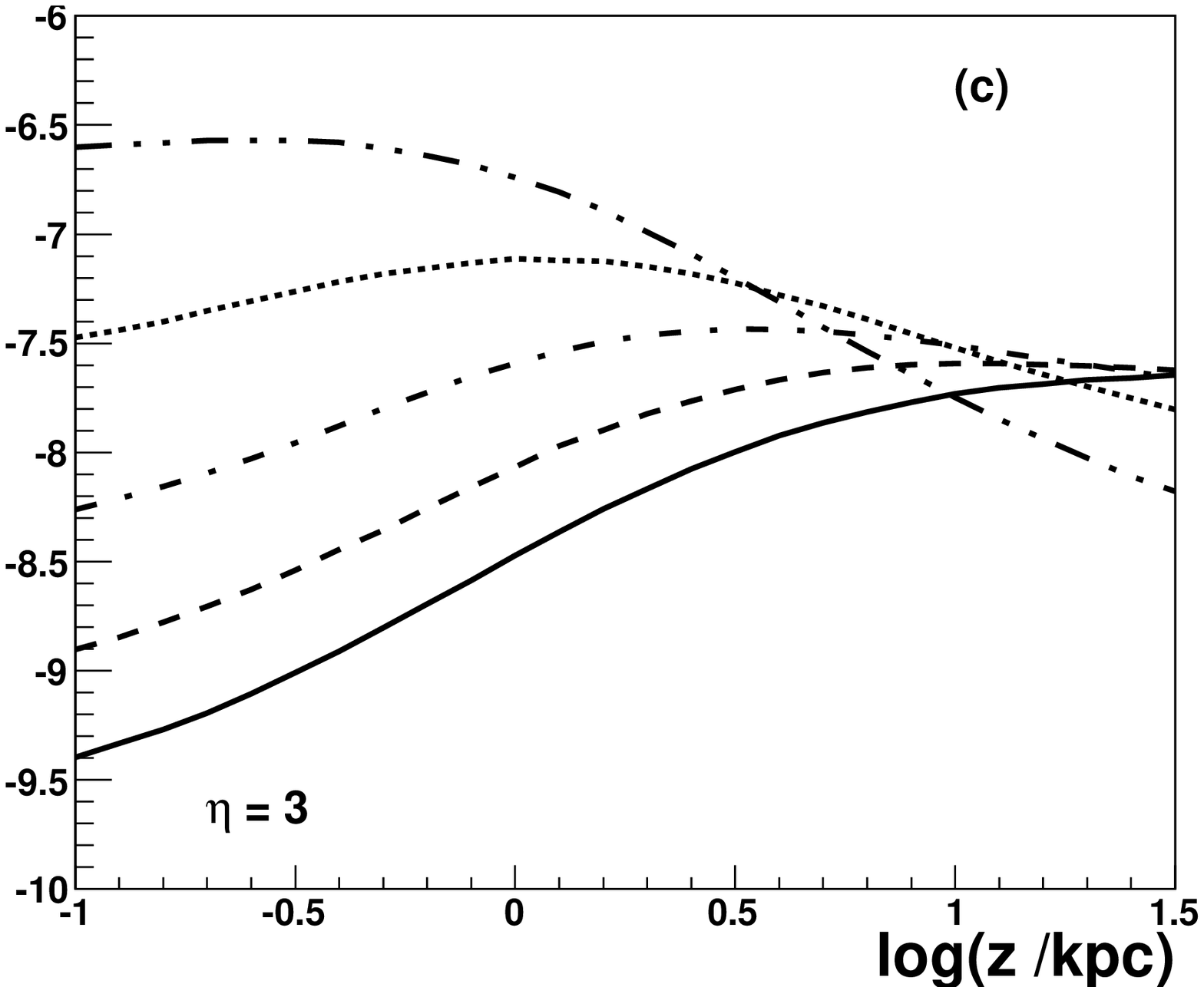}
\includegraphics{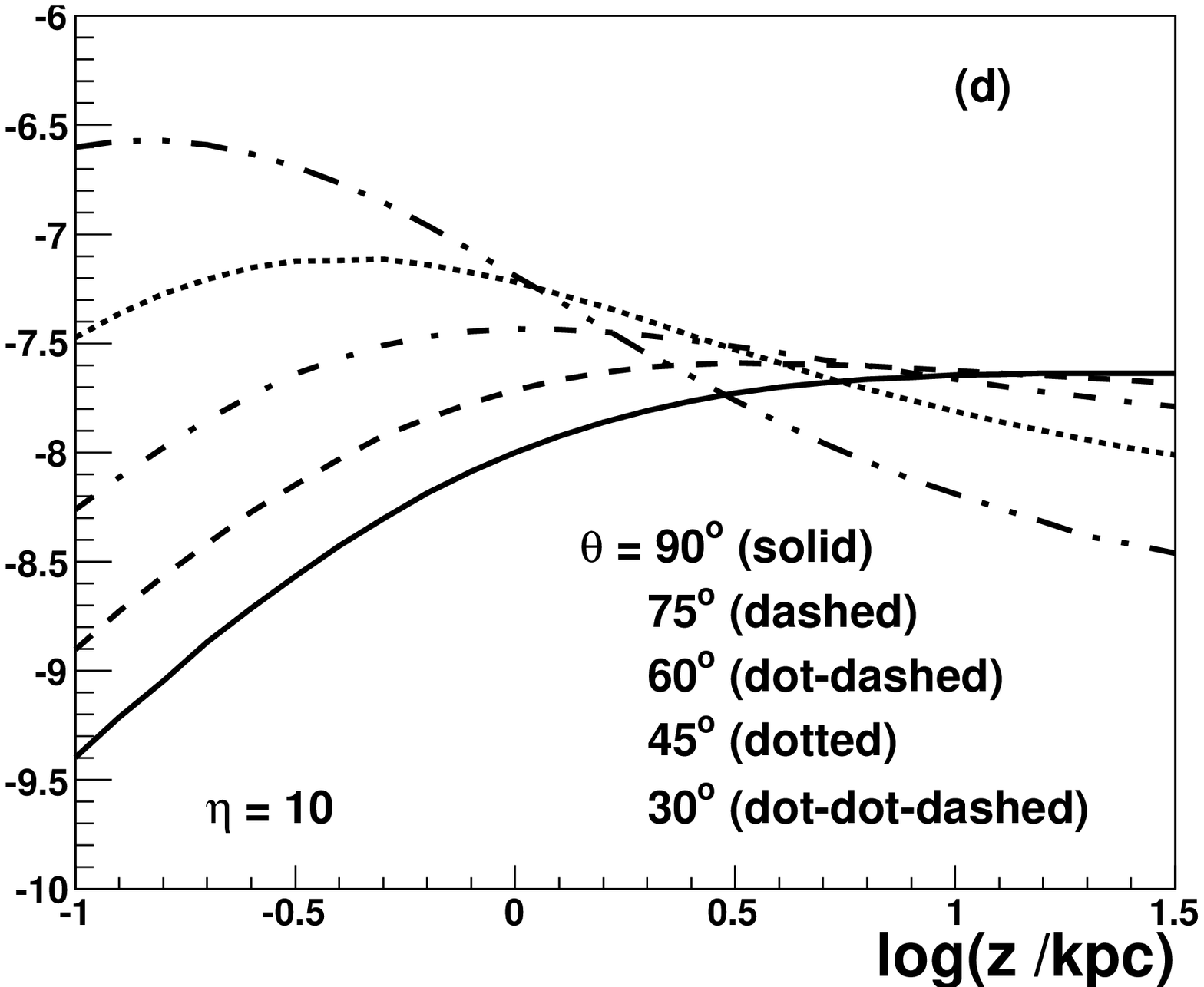}
\caption{The $\gamma$-ray flux ($>$300 GeV), as a function of the real distance from the base of the jet. It is produced for different structure of the jet described by the parameter $\eta = 0.3$ (figure a), 1 (b), 3 (c), and 10 (d), in the case of injection of relativistic electrons with a power law spectrum (spectral index 2 extending between 1 GeV and 30 TeV and normalized to 1 MeV~s$^{-1}$~kpc$^{-1}$). The electrons cool locally in the jet as a result of the IC scattering of the soft radiation which is produced in the inner jet with the spectrum observed from Cen~A. The synchrotron energy losses of the electrons are neglected. Specific curves show the fluxes for the observation angles 
$\theta = 90^\circ$ (solid), $75^\circ$ (dashed), $60^\circ$ (dot-dashed), $45^\circ$ (dotted), and $30^\circ$ (dot-dot-dashed). The other parameters of the model are $\gamma_{\rm i} = 20$, $\gamma_{\rm o} = 3$, and $z_{\rm 0} = 0.1$~kpc.} 
\label{fig8}
\end{figure*}

At first, we investigate the morphology of the $\gamma$-ray production from jets in radio galaxies (i.e. the dependence of $\gamma$-ray flux above 300 GeV on the distance along the jet) assuming constant injection rate of the electrons along the jet. Different models for the velocity structure of the jet and its inclination to the observer's line of sight are considered (see Fig.~8). The distribution of the 
$\gamma$-ray flux along the jet shows interesting dependence. It starts to change their behavior at distances for which the jet significantly decelerates. This happens at distances of a few kpc for slowly decelerating jets (Fig.~8a), but already at $\sim$0.1 kpc for  fast decelerating jets (Fig.~8c). The morphology of the $\gamma$-ray emission (its dependence on the distance and the angle $\theta$) is not easy to understand. It is determined by a few different factors which affects the fluxes in the opposite directions. The most important are: (1) the $\gamma$-ray emission pattern in the electron reference frame; (2) the density of the soft radiation in the kpc scale jet; and (3) the angular dependence of $\gamma$ rays on the Lorentz factor of the jet. All those factors depend on the distance from the base of the jet and on the observation angle $\theta$. In general, at small distances from the base of the jet, the $\gamma$-ray fluxes are the largest for small angles $\theta$. However, at large distances from the base of the jet, the $\gamma$-ray fluxes are on a similar level, independent on the observation angle. The change of the $\gamma$-ray emission pattern appears at lower distances from the base of the jet in the case of fast decelerated jets (see Fig.~8d).    

Note that, the example $\gamma$-ray morphology from the kpc scale jet (as shown in Fig.~8) are obtained under the assumption that the injection rate of the relativistic electrons is constant with the distance from the base of the jet. This is probably idealistic assumption in the case of the kpc scale jets of radio galaxies. However, it allows us to understand the basic features of the considered model. Unfortunately, the injection rate of electrons (and their spectrum) is not at present well constrained by the observations. In fact, the $\gamma$-ray emission morphology (in Figs.~8) should be convolved with the efficiency of electron acceleration in the realistic jets. The problem may be reversed with the future observations with the next generation instrument such as CTA.
Then, the comparison of the distribution of $\gamma$-ray emission, along the jet of a  specific radio galaxy (e.g. Cen~A or M87) with the considered here model should allow us to constrain the efficiency of the electron injection rate into the kpc scale jet and also the parameters of the jet defined in our simple case by $\gamma_{\rm 0}$, $z_{\rm 0}$, and $\eta$.     

\begin{figure}
\vskip 15.truecm
\includegraphics{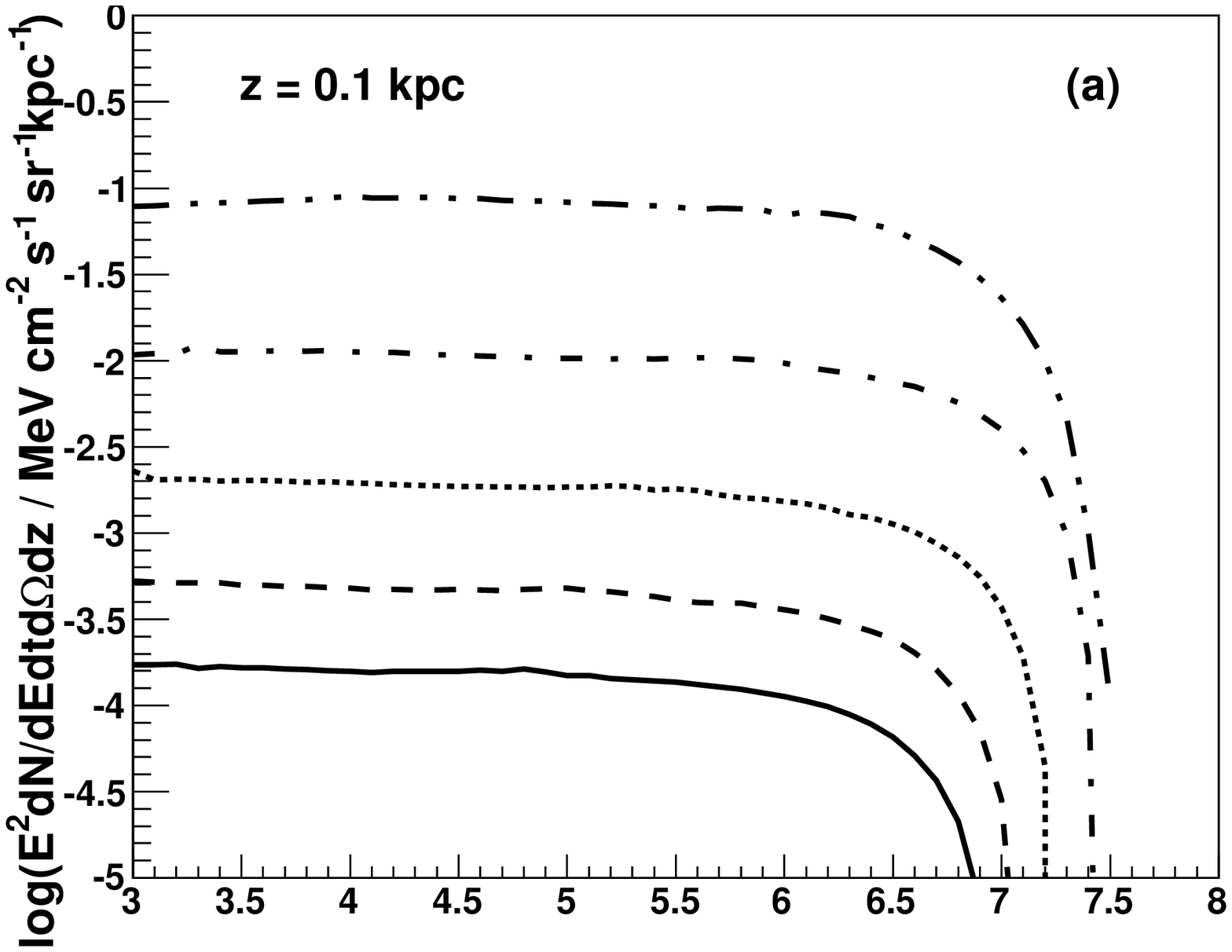}
\includegraphics{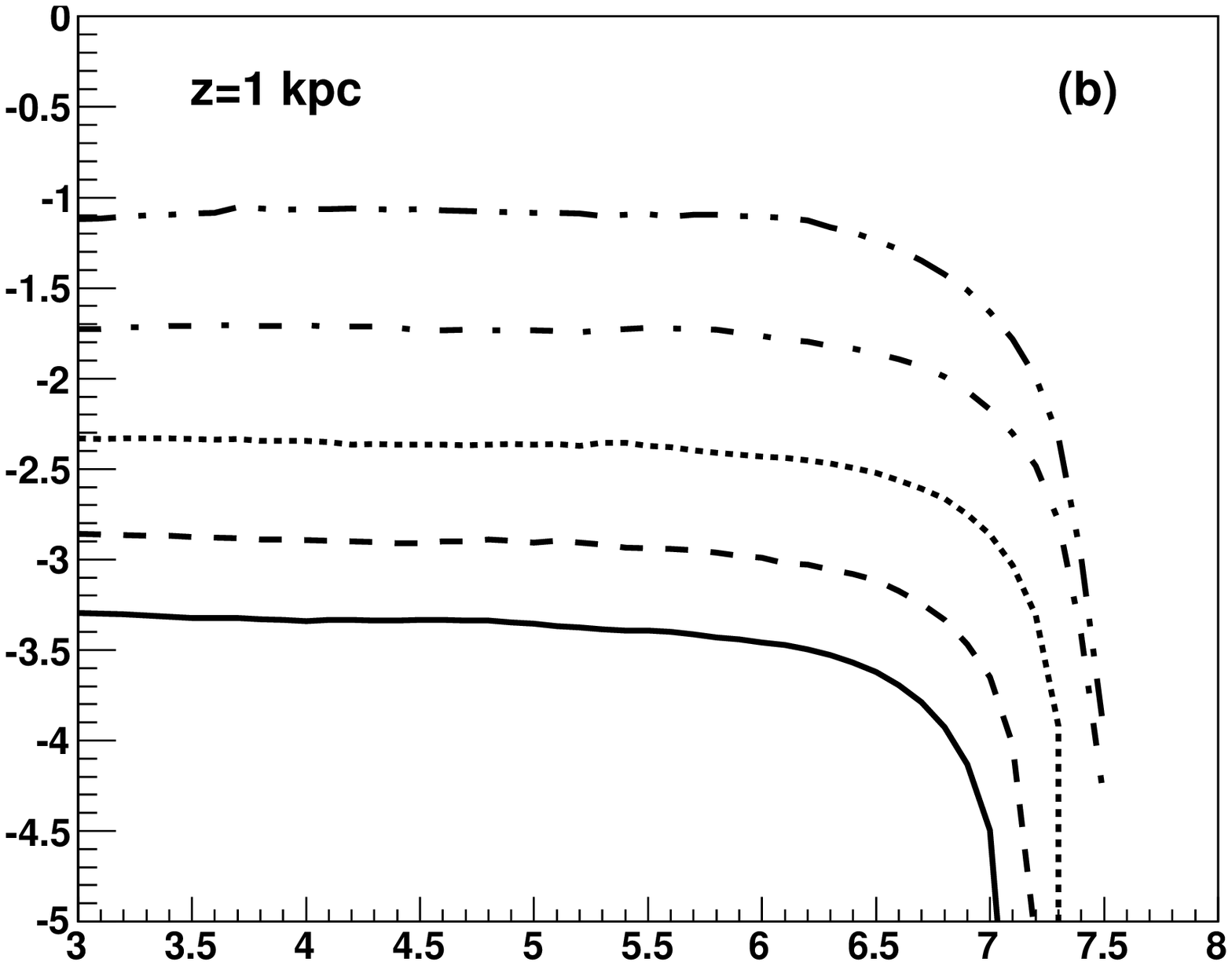}
\includegraphics{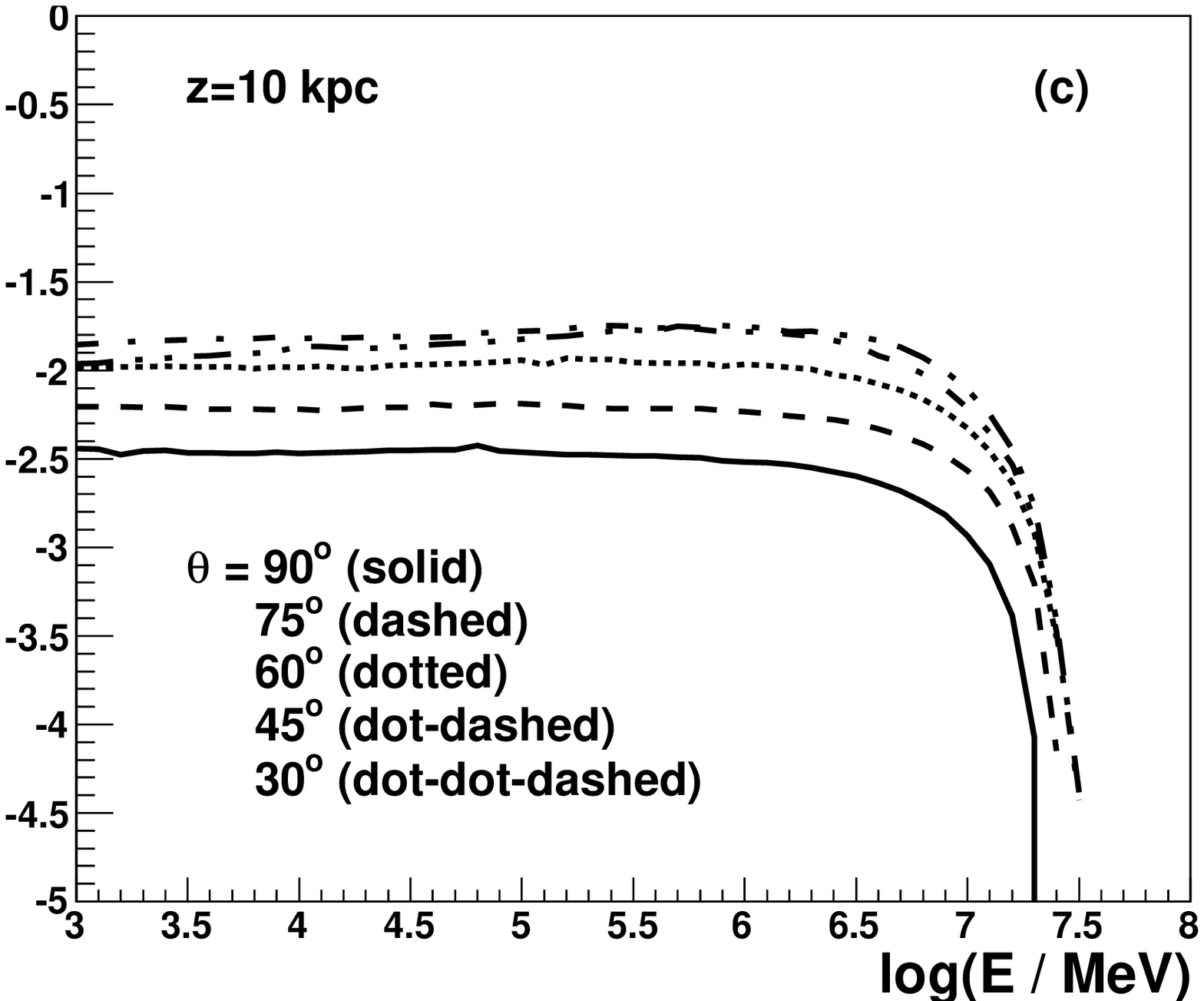}
\caption{The example SED of $\gamma$-rays produced  for the parameters as indicated in Fig.~8, at different distances from the base of the jet $z = 0.1$ kpc (figure a), 1 kpc (b), and 10 kpc (c). The spectra are shown for different inclination angles of the jet $\theta = 30^\circ$ (dot-dot-dashed), 
$45^\circ$ (dot-dashed), $60^\circ$ (dotted), $75^\circ$ (dashed), and $90^\circ$ (solid). The velocity structure of the jet is described by the parameter $\eta =1$. The other parameters of the model are 
$\gamma_{\rm o} = 3$ at $z_{\rm o} = 0.1$~kpc.} 
\label{fig9}
\end{figure}

We also show the $\gamma$-ray spectra, produced at different distances from the base of the jet, in the case of negligible synchrotron energy losses of the electrons, for the power law spectrum of electrons with the index equal to 2 (see Fig.~9). The shapes of the $\gamma$-ray spectra are very similar at different distances from the base of the jet and for different inclination angles. The spectra have also the power law type, with the spectral index close to 2. This is due to their dominant IC cooling process. These $\gamma$-ray spectra show steepening at the largest energies due to the maximum energies of injected electrons. 

In realistic jets, the electrons can suffer energy losses not only on the IC process but also on the synchrotron radiation. Therefore, we investigate the modification of the $\gamma$-ray fluxes under the additional influence of the synchrotron energy losses. The example calculation of the $\gamma$-ray emission for the case of the synchrotron energy losses of electrons in the kpc scale jet are investigated in Fig.~10. Note that, for the specific parameters (mentioned in Fig.~10), $\gamma$-ray fluxes start to be affected at the kpc scale distances if the magnetic field strength becomes of the order of a few $10^{-4}$~G. Such order of magnitude magnetic fields are expected within the kpc scale jets of the radio galaxies (e.g. Pudritz, Hardcastle~\& Gabuzda~2012). Therefore, we conclude that when investigating the geometrical structure of the $\gamma$-ray emission from the kpc scale jets, the energy losses of the electrons on the synchrotron process can effect the production of $\gamma$ rays in terms of the considered model at the kpc scale distances from the base of the jet. The observations of the $\gamma$-ray fluxes from kpc scale jets of specific radio galaxies, together with their radio to X-ray synchrotron emission, should allow us to put constraints on the structure of the magnetic field along  their kpc scale jets.  

\begin{figure}[t]
\vskip 5.5truecm
\includegraphics{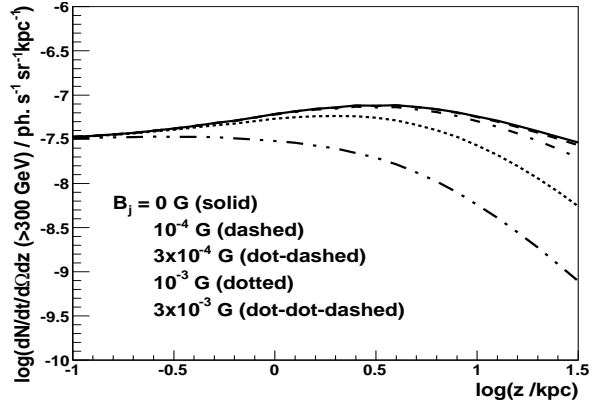}
\caption{The distribution of the $\gamma$-ray flux in the case of different strength of the magnetic field within the jet, $B_{\rm j} = 0$ G (solid curve), $10^{-4}$~G (dashed), $3\times 10^{-4}$~G (dot-dashed), $10^{-3}$~G (dotted), and $3\times 10^{-3}$~G(dot-dot-dashed). The other parameters of the model are  $\theta = 45^\circ$, $\gamma_{\rm i} = 20$, $\gamma_{\rm o} = 3$, $\eta = 1$,
and $z_{\rm 0} = 0.1$~kpc. Electrons are injected with the power law spectrum (spectral index 
$\alpha = 2$ between 1 GeV and 30 TeV, normalized to 1~MeV~s$^{-1}$~kpc$^{-1}$). They cool locally in the jet due to energy losses on the IC scattering of the radiation from the inner jet (the spectrum of Cen~A is assumed) and on the synchrotron process in the magnetic field. Other parameters of the model are the same as in Fig.~8.}
\label{fig10}
\end{figure}
\subsection{Gamma-ray spectra from the whole jet}

In practice, it is difficult to study details of the morphology of the $\gamma$-ray emission even from the nearby radio galaxies due to insufficient angular resolution of the present Cherenkov telescopes. Therefore, we also calculate the expected $\gamma$-ray fluxes from the whole kpc scale jets. As above, the electrons are injected into the jet with the power low spectrum defined by the spectral index equal to 2. We now relax previous assumption on the constant injection rate of the electrons along the jet. We compare the constant injection rate model 1, i.e. $A_{\rm z} = A_{\rm 0} =1$~kpc$^{-1}$, with the models in which the power law dependence of electron injection rate is considered  according to model 2: $A_{\rm z} = A_1 z^{-1}$~kpc$^{-1}$; and model 3: $A_{\rm z} = A_2 z^{-2}$~kpc$^{-1}$. The normalization  $A_1$ and $A_2$ are obtained from the condition $\int_{z_{\rm min}}^{z_{\rm max}}A_{\rm z} dz = 1$ in the range of distances between $z_{\rm min} = 0.1$ kpc and $z_{\rm max} = 30$ kpc. These constants are then equal to $A_{\rm 1} = [\ln(z_{\rm max}/z_{\rm min}]^{-1}$ and 
$A_{\rm 2} = z_{\rm min}z_{\rm max}/(z_{\rm max}-z_{\rm min})$~kpc.

The example total $\gamma$-ray fluxes ($>$300~GeV), integrated over the range of distances from 0.1 kpc to 30 kpc, are shown as a function of the inclination angle of the jet in Fig.~11.  
It is evident from these calculations that for the observation angles lower than several degrees the 
$\gamma$-ray fluxes drop abruptly in all models due to the kinematics of the IC scattering process. The largest fluxes are predicted for the range of the inclination angles between $\sim$10$^\circ$ to 
30$^\circ$, provided that jets do not decelerate fast. The fast decelerated jets (i.e. the parameter $\eta$ large) produce dominant $\gamma$-ray fluxes at large distances from the base of the jet. 
We note generally declining tendency of the $\gamma$-ray fluxes with the distance from the base of the jet. It is due to the weakening of the soft radiation field from the inner jet and also the assumed shapes of the functions describing the injection rate of the electrons. The above mentioned effects are the strongest in the case of weakly decelerated jets since such jets move with significant Lorentz factors at large distances from the base of the jet.

\begin{figure}
\vskip 15.5truecm
\includegraphics{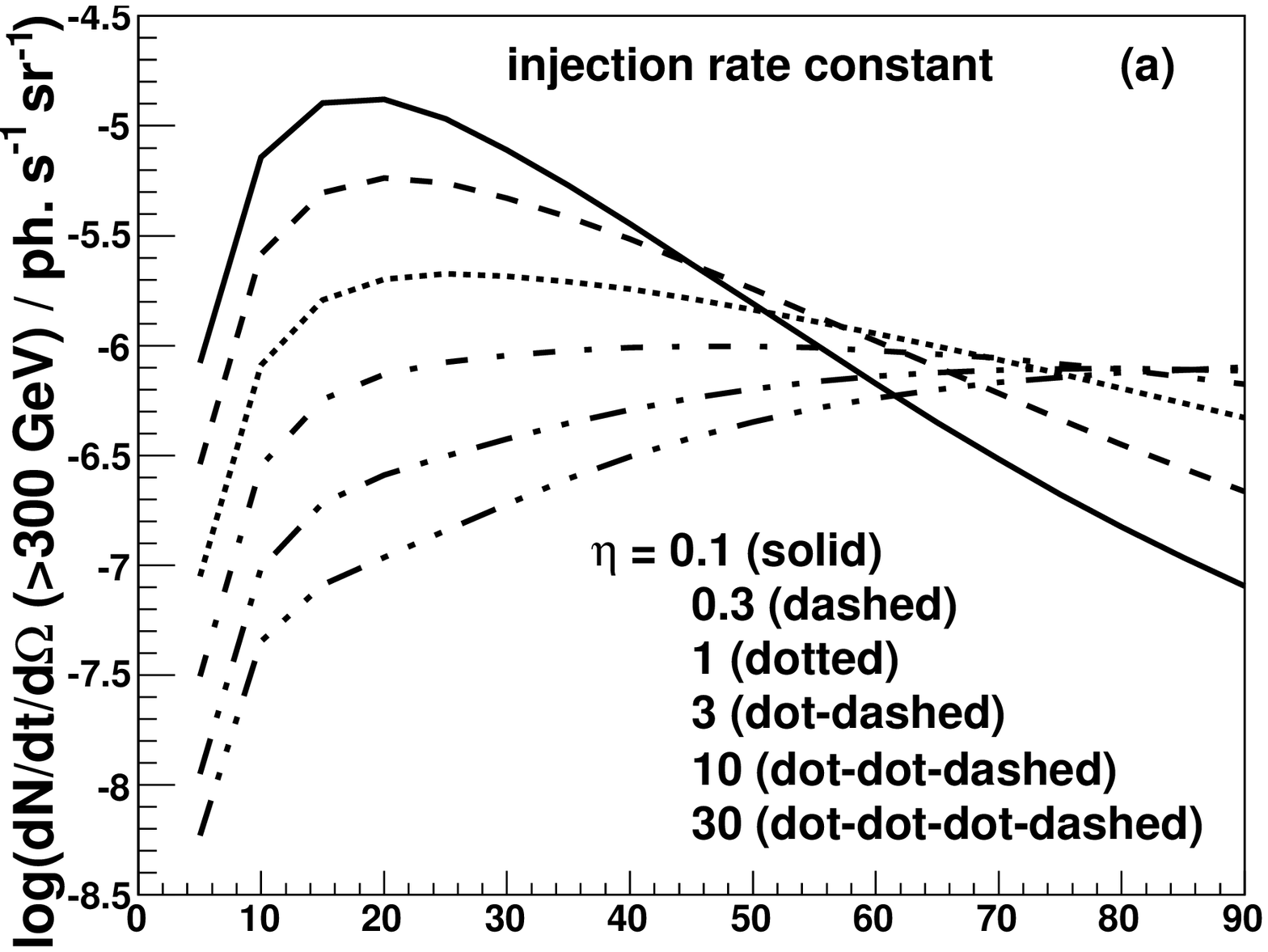}
\includegraphics{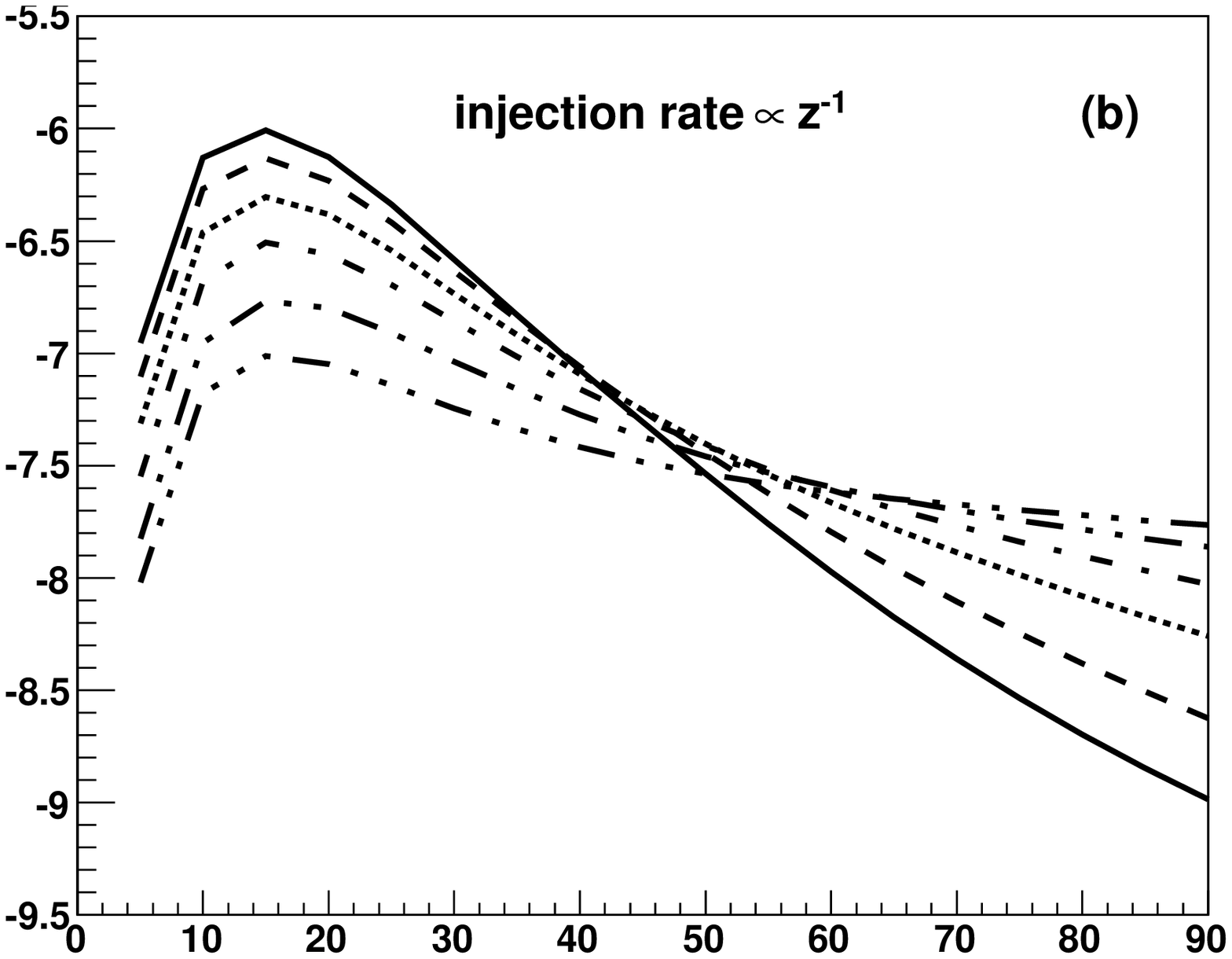}
\includegraphics{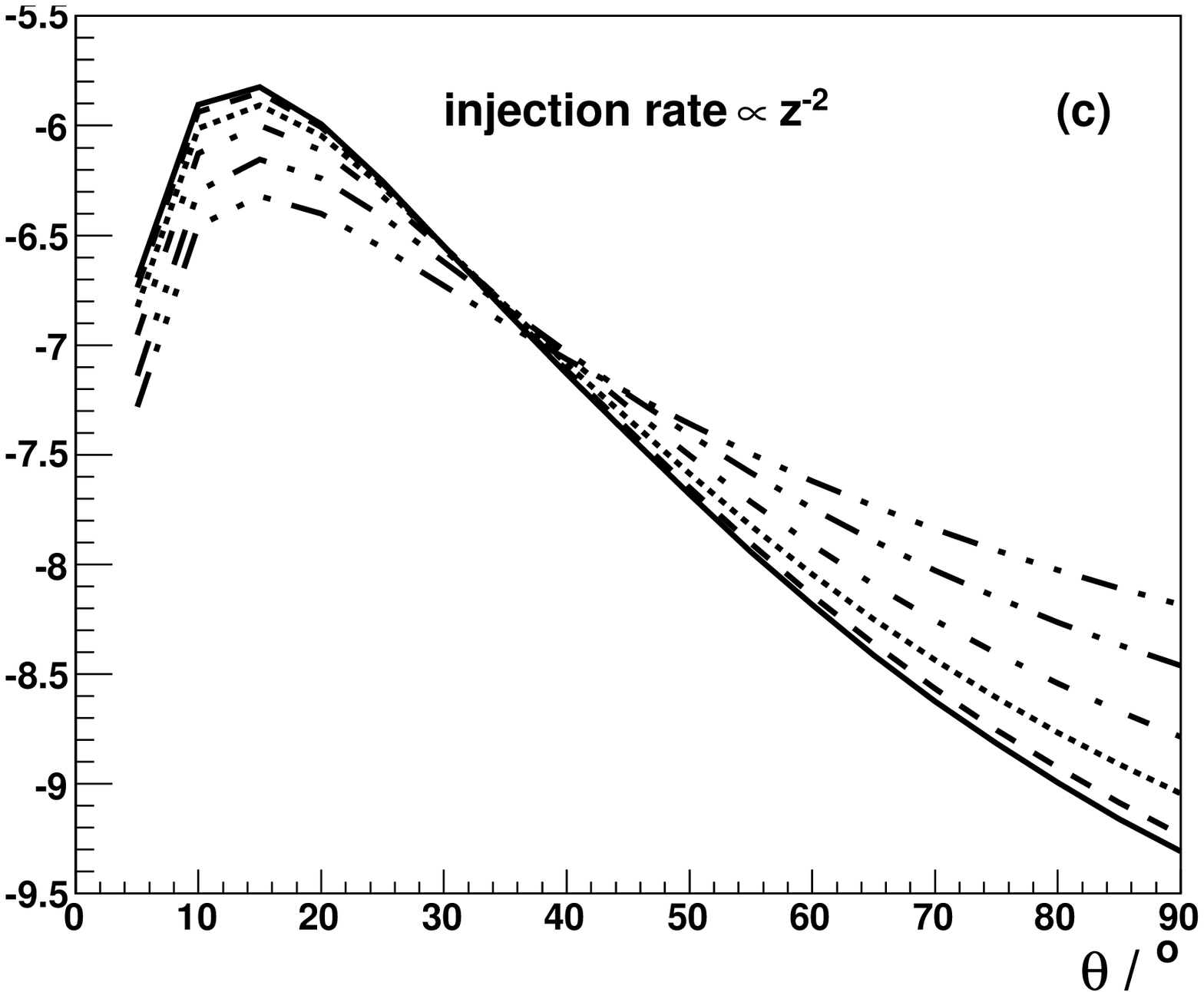}
\caption{The $\gamma$-ray flux produced in the whole jet (between $z_{\rm min} = 0.1$ kpc and $z_{\rm max} = 30$ kpc), assuming that electrons are injected along the jet with a constant rate and their spectrum is described by the power law function (given by Eq.~6). This spectrum has the spectral index $\alpha = 2$. The angular dependence of the flux is shown for different models of the jet velocity structure described by $\eta = 0.1$ (solid), 0.3 (dashed), 1 (dotted), 3 (dot-dashed), 10 (dot-dot-dashed), and 30 (dot-dot-dot-dashed). The other parameters of the model are $\gamma_{\rm o} = 3$, $z_{\rm o} = 0.1$~kpc, and $\gamma_{\rm i} = 20$. Specific figures show the fluxes for different dependence of the injection rate of the electrons on the distance from the base of the jet defined by the models (1), (2) and (3) as shown in Sect.~5.3.} 
\label{fig11}
\end{figure}

We also investigated the dependence of the $\gamma$-ray fluxes ($>$300~GeV) on the initial Lorentz factor of the jet (measured at the distance $z_{\rm 0} = 0.1$~kpc). As expected, for large initial Lorentz factors, the $\gamma$-ray fluxes drop strongly with the observation angle $\theta$ in all considered models for the electron injection rate. This effect is stronger for the fast decelerated jets (see dot-dashed, thin and thick curves in Fig.~12). In such a case, the $\gamma$-ray emission preferentially dominates close to the base of the jet, provided that it is  seen at small observation angle. On the contrary, initially slow jets emit $\gamma$-rays mainly at large observation angles. For example, $\gamma$-ray flux at $\sim$10$^\circ$ becomes 2-3 orders of magnitude lower than the flux at the angles above $\sim$30$^\circ$. 

The $\gamma$-ray spectra expected from the whole jet, in the considered range of distances, are presented in Fig.~13. The angular distribution of the $\gamma$-ray fluxes, for the case of a constant injection rate of the electrons as a function of the distance from the jet base, are shown in Fig.~11a. We also investigate dependence of the $\gamma$-ray spectra on different models for the velocity structure of the jet (described by the deceleration parameter $\eta$) and their observation angle $\theta$. 
As expected in the case of the complete local cooling of the electrons injected with a power low spectrum and spectral index 2, the $\gamma$-ray spectra are also characterized by the spectral index close to 2. The spectra, in the case of fast decelerated jets, show the maximum at slightly lower energies for small angles 
$\theta$ (see e.g. Fig.~13a for $\theta = 30^\circ$). They are also clearly flatter than the spectra for slowly decelerated jets. On the other hand, for large angles $\theta$ (see Fig.~13b for $\theta = 90^\circ$),
the $\gamma$-ray spectra from the fast decelerated jets (triple-dot-dashed curves) are slightly harder. They also extend to larger energies. These tiny effects are due to the fact that fast decelerated jets move through the kpc scale distances with the Lorentz factors which are not far from unity. In such a case, the angular pattern of the 
$\gamma$-ray spectra, produced by the electrons from the blob that is almost at rest, plays the dominant role. Finally, the dependence of the 
$\gamma$-ray spectra on the observation angle $\theta$ is investigated for the fixed model of the velocity structure of the jet (Fig.~13c). In such a case, the spectra are almost independent on the observation angle. Note however, a small flattening of the $\gamma$-ray spectrum  for the intermediate angles $\theta$ (dot-dot-dashed curve in Fig.~13c). 

The spectral indexes of the TeV $\gamma$-ray emission from two best studied radio galaxies are steeper than 2. For example, Cen~A shows the $\gamma$-ray spectrum in the TeV energies well described by the spectral index close to 2.5 (Abdalla et al.~2018), and the spectral index at TeV energies from M~87 in the low state is close to 2.4 (e.g. Acciari et al.~2020). 
Although, it is not clear whether all that emission is produced in the intermediate scale jet, it
is possible that the spectral index of the TeV $\gamma$-ray emission from the kpc scale jet is also steeper than 2. Due of those uncertainties, we also show the $\gamma$-ray spectra calculated for the range of spectral indexes of the electrons injected into the kpc scale jet (see upper panel in Fig.~14). The spectral indexes
of this $\gamma$-ray emission are closely related to the spectral indexes of the spectra of the injected electrons. In the case of a complete and local cooling of the electrons, the spectral index of the equilibrium spectrum of the electrons should be steeper by unity. On the other hand, the electrons with the equilibrium spectrum should produce $\gamma$-rays in the Thomson regime with the spectral index close to $(\alpha + 2)/2$, i.e. for the injection spectrum of the electrons equal to 3, the observed spectrum of the $\gamma$-rays should have the spectral index close to 2.5.  Therefore, we conclude that the observed spectral index of the $\gamma$-ray emission (close to 2.5) requires the injection of the electrons into the kpc scale jet with the spectral index close to 3. As an example, we calculate the morphology of the $\gamma$-ray emission in the case of the electrons injected with the spectral index equal to 3 (see bottom panel in Fig.~14). It is very similar to the morphology expected for the electrons injected with the index equal to 2 (see Fig.~8d). Therefore, we conclude that the morphology of the $\gamma$-ray emission is independent on the spectral index of the injected electrons but it is determined by the parameters defining the evolution of the jet and its observation angle as indicated in Fig.~8.

\begin{figure}
\vskip 14.5truecm
\includegraphics{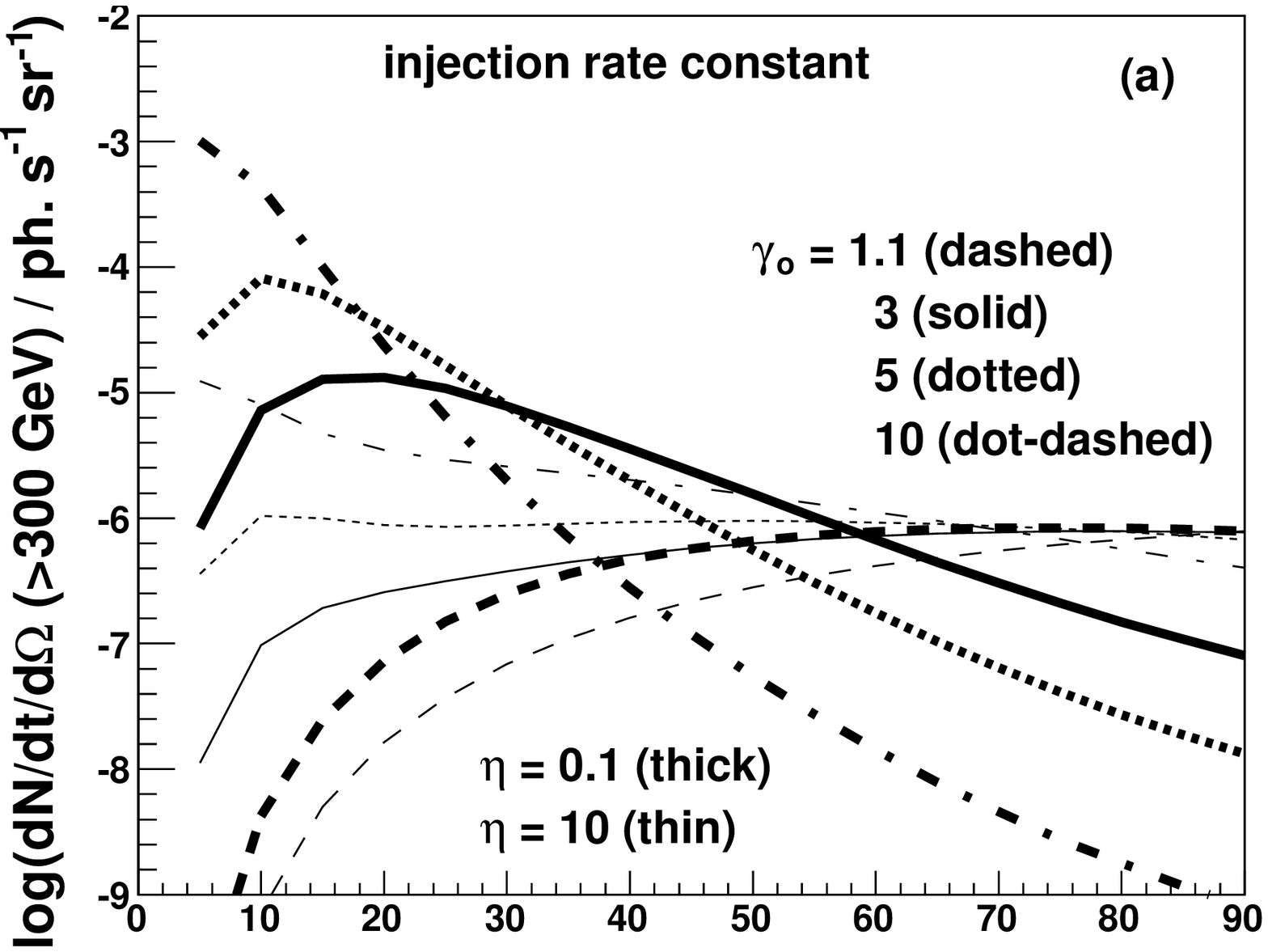}
\includegraphics{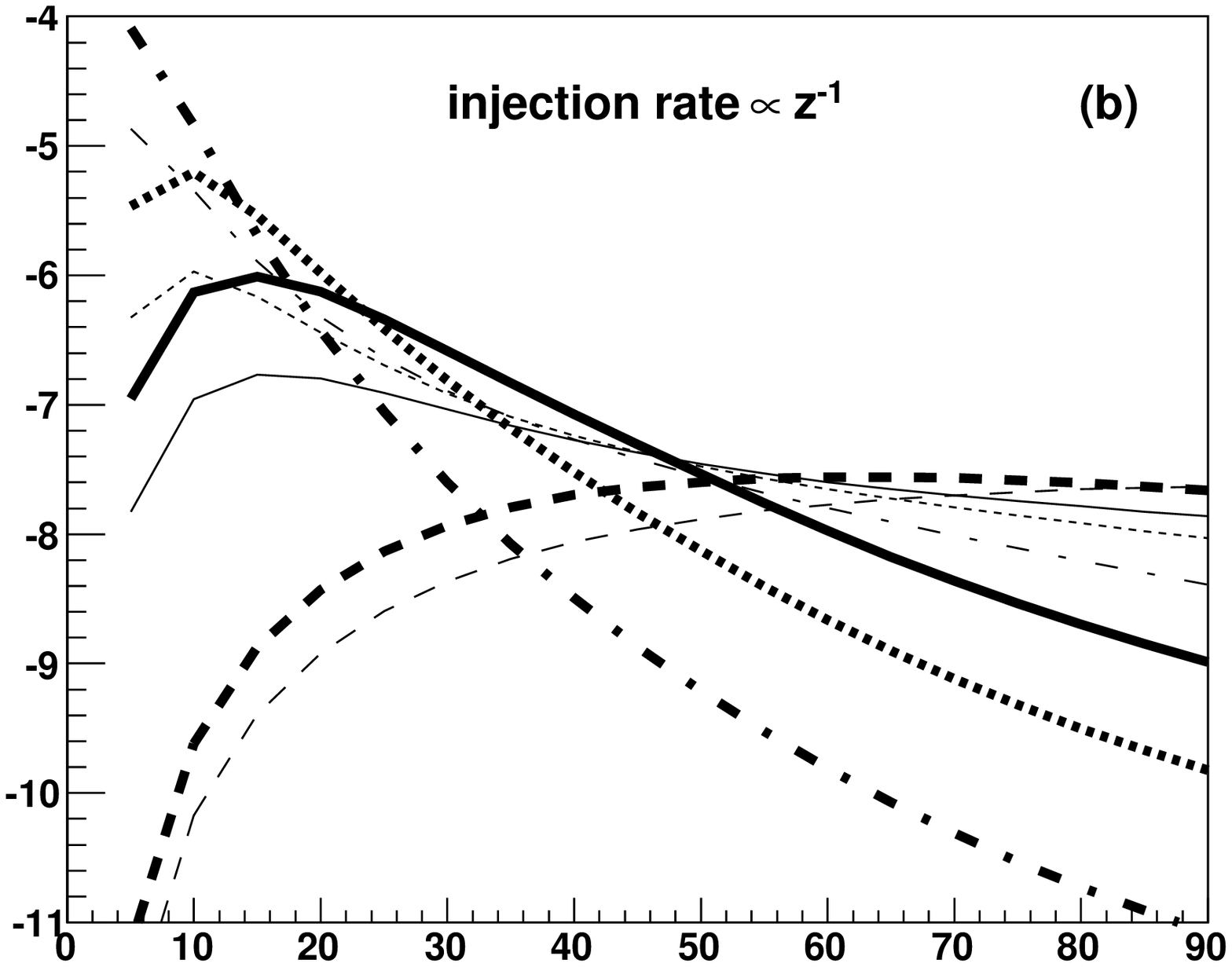}
\includegraphics{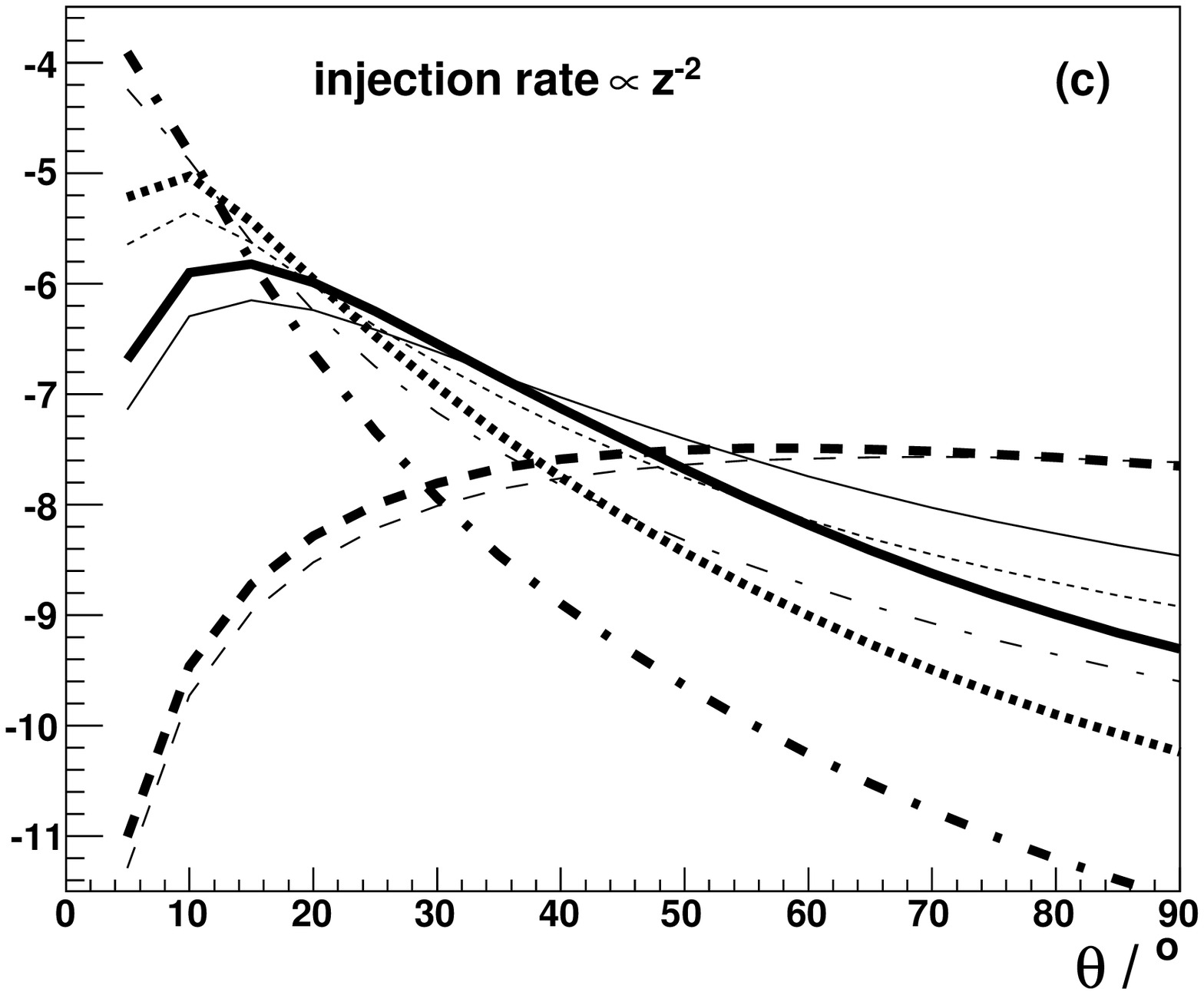}
\caption{The $\gamma$-ray flux (as in Fig.~11) but for the models for the velocity structure of the kpc scale jet described by different values of the initial Lorentz factor of the jet at $z_{\rm 0} = 0.1$~kpc, equal to $\gamma_{\rm 0} = 1.1$ (dashed curve), 3 (solid), 5 (dotted) and 10 (dot-dashed) for $\eta = 0.1$ (thick curves) and $\eta = 10$ (thin curves). The other parameters describing these models are as in Fig.~11.} 
\label{fig12}
\end{figure}
\section{Discussion and conclusions}

Following general unified model for the high energy radiation in active galaxies  (e.g. Barthel~1989, Urry \& Padovani~1995), it is believed that radio galaxies are BL Lac type objects viewed at a relatively large inclination angles in respect to their jet direction. Therefore, we assumed that the inner jets in radio galaxies also move relativistically with the Lorentz factors typically of the order of 10-30. It is shown (Bednarek 2019) that in such a case the soft radiation background at kpc scale distances from the base of the jets is likely dominated by the non-thermal inner jet emission (in case of straight jets) but not by other radiation fields, such as e.g. the starlight from the host galaxy, the cosmic microwave background, or from the accretion disk and the molecular torus (see also  Tanada et al.~2019). Then, $\gamma$-rays, produced in the IC scattering of this soft radiation by relativistic electrons distributed isotropically in the kpc scale jet reference frame, should have specific spectral proprieties. Due to the geometry of the IC process, defined by almost mono-directional soft radiation field along the jet and the isotropic relativistic electrons at the kpc distance scale along the jet, dominant $\gamma$-ray emission is also expected at large angles to the jet axis. We study in detail the proprieties of such TeV $\gamma$-ray emission for different jet models and proprieties of the injected electrons. We show that the electrons, injected with a power law spectrum into the jet, cool locally in the jet on the synchrotron and the IC processes for likely conditions in the jet (see Fig.~7). In such a case, we determine the local equilibrium spectrum of the electrons in the jet.  For these equilibrium electron spectra, we calculate the GeV-TeV $\gamma$-ray spectra for different models for the deceleration of the jet and inclination angles of the jet in respect to the direction towards the observer. 
It is shown that fast and slowly decelerated jets show different dependence of the $\gamma$-ray emission on the distance from the base of the jet (Fig.~8). Moreover, the $\gamma$-ray flux also depends on the observation angle of the jet. We also consider jets in which the electrons are injected with a different rate along the jet. The possible role of the synchrotron energy losses of the electrons within the kpc scale jet is also considered. We show that the magnetic fields with the strengths above $\sim$10$^{-4}$~G should  already influence the expected TeV $\gamma$-ray emission from the kpc scale jets (see Fig.~10). In fact, magnetic field with strengths of a similar magnitude are reported from the kpc scale jets in active galaxies (see e.g. review by Pudritz et al.~2012 and references therein).    

\begin{figure}
\vskip 16.truecm
\includegraphics{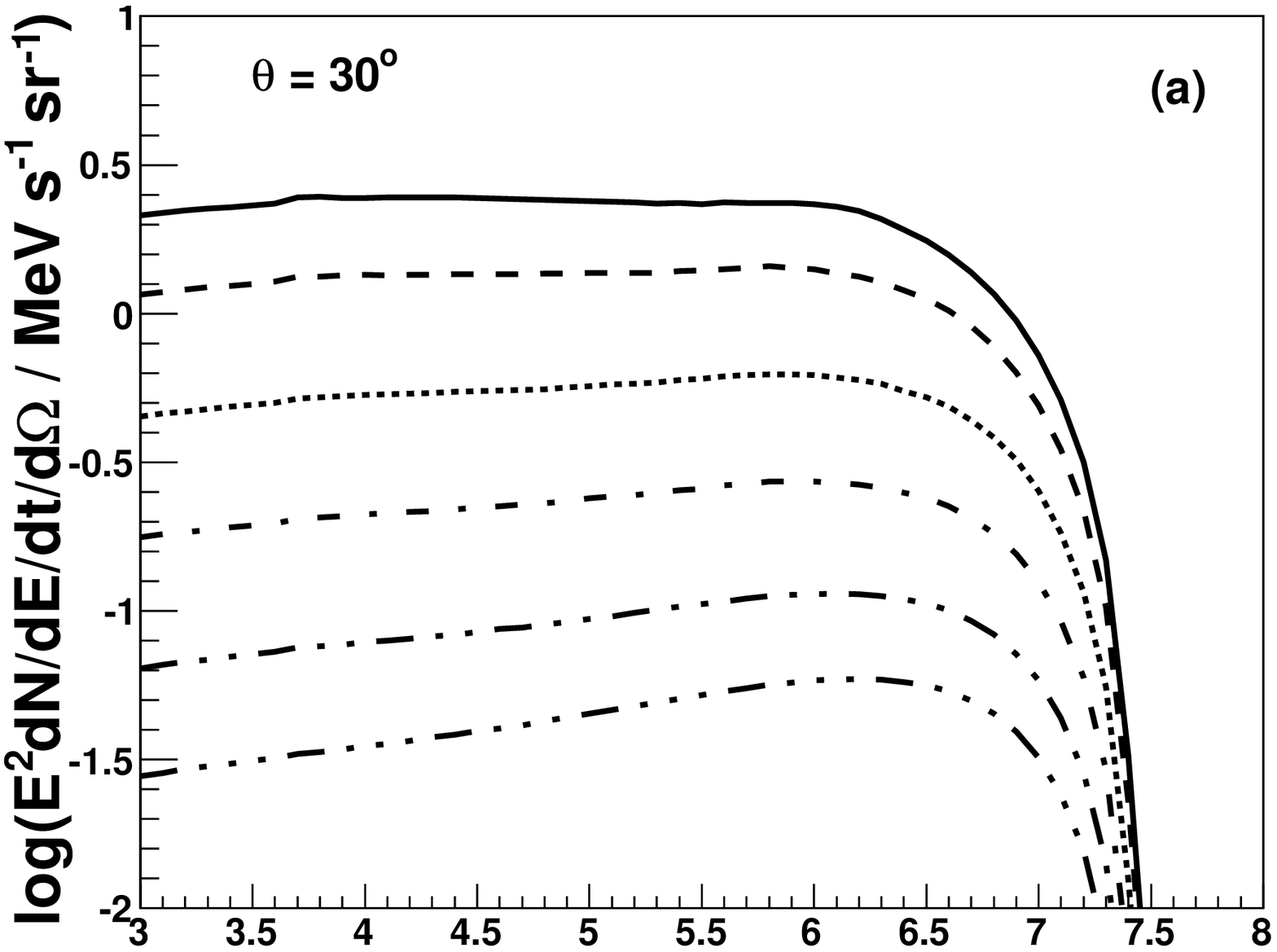}
\includegraphics{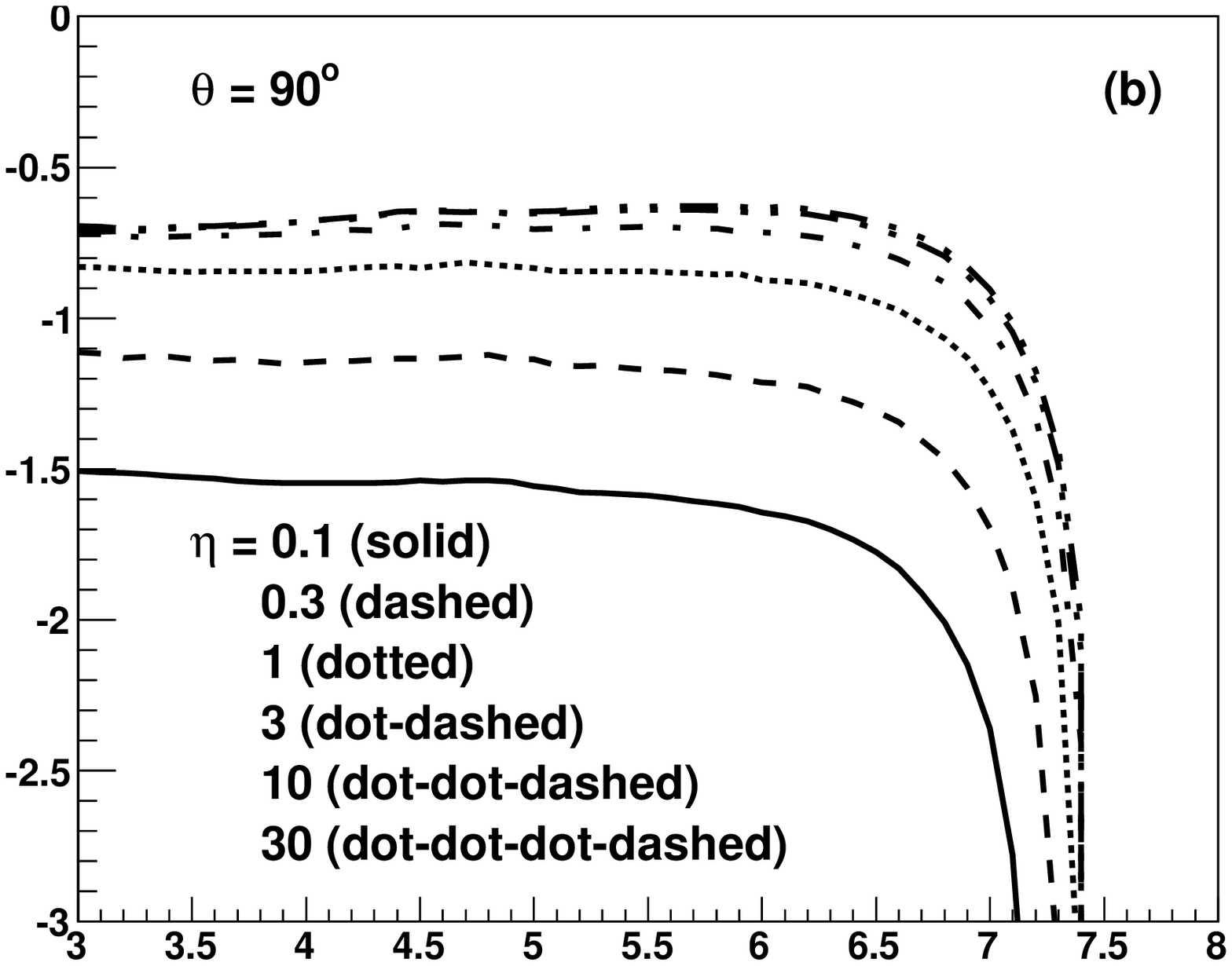}
\includegraphics{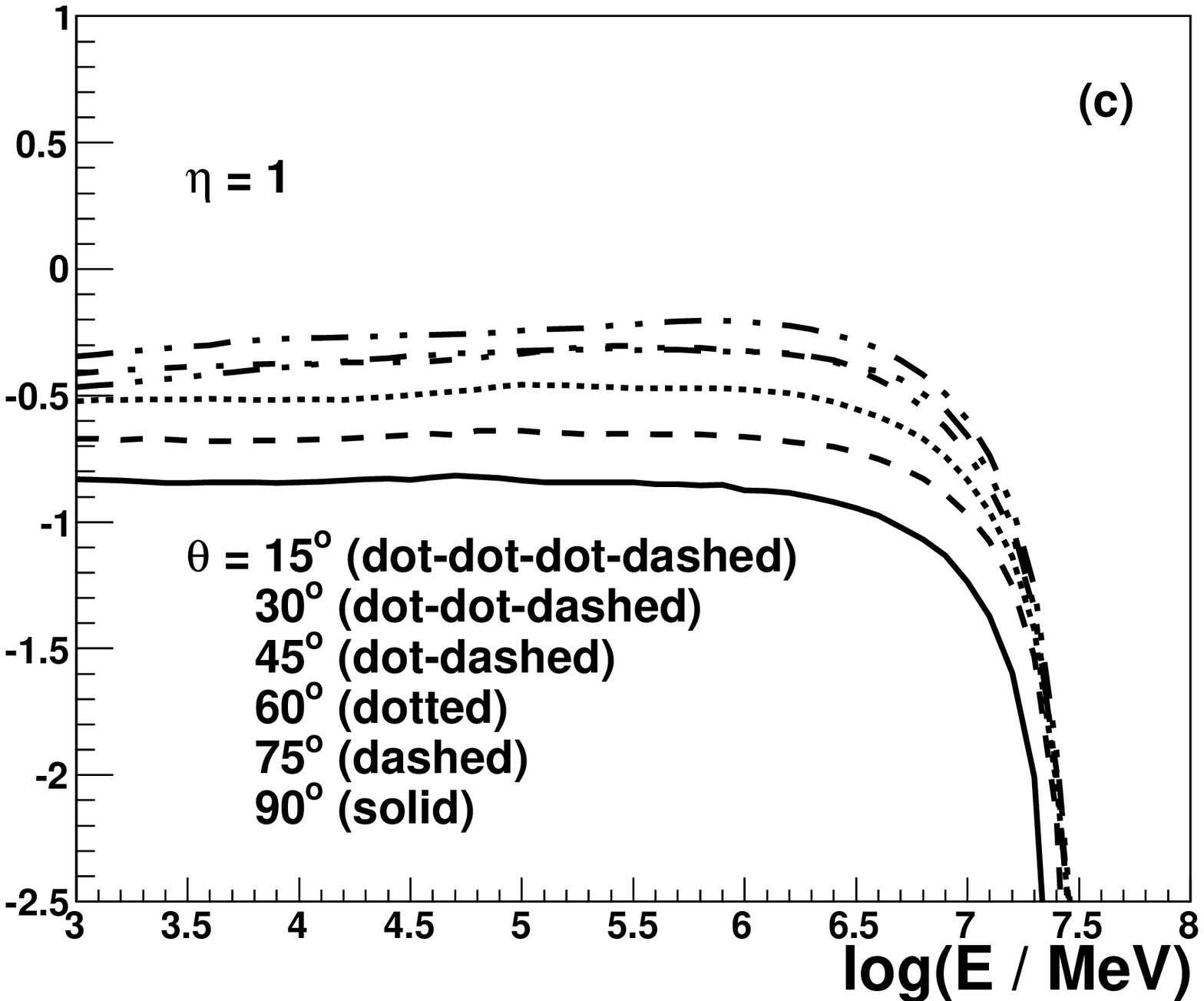}
\caption{The example SEDs of $\gamma$-rays from the kpc scale jet, integrated over the range of distances between $z = 0.1$ kpc and 30 kpc, are shown for different models of the jet velocity structure $\eta = 0.1$ (solid), 0.3 (dashed), 1 (dotted), 3 (dot-dashed), 10 (dot-dot-dashed), and 30 (dot-dot-dot-dashed), and for the inclination angles of the jet $\theta = 30^\circ$ (figure a) and $90^\circ$ (b). 
The SEDs, for selected angles $\theta = 15^\circ$ (dot-dot-dot-dashed), $30^\circ$ (dot-dot-dashed), $45^\circ$ (dot-dashed), $60^\circ$ (dotted), $75^\circ$ (dashed), and $90^\circ$ (solid) and fixed value of $\eta = 1$, are shown in (c). Other parameters of the model are $\gamma_{\rm i} = 20$, $\gamma_{\rm o} = 3$, $z_{\rm 0} = 0.1$~kpc, and $A_{\rm z} = 1$. The complete local cooling of the electrons, injected with a power law spectrum and spectral index 2 up to 30 TeV, is assumed.} 
\label{fig13}
\end{figure}

We conclude that the kpc scale jets should produce $\gamma$-rays efficiently  at large inclination angles to the jet direction. However, the efficiency of the $\gamma$-ray emission depends on a few important parameters which describe the injection of the electrons and the dynamics of the jet at kpc scale distances. They can be constrained by direct comparisons of the model predictions, such as discussed here, with the observations of specific radio galaxies. Such comparisons are still very difficult in the
$\gamma$-ray energy range even in the case of the nearby radio galaxies due to the limited angular resolution of the present Cherenkov telescopes. However, first detection of the extended TeV $\gamma$-ray emission from the radio galaxy Cen~A has been recently reported (Sanchez et al.~2018), proving that such an emission is produced in the kpc scale jets.  This TeV $\gamma$-ray emission has been found to originate at the projected distance of $R_{\rm p} \sim$3~kpc from the base of the jet in Cen~A (Sanchez et al.~2018). In fact, the real distance from the base of the jet must be even larger ($R=R_{\rm p}/\sin\theta$) due to the projection effect. For the inclination angle of the jet, as estimated in Cen~A in the range $(12 - 80)^\circ$ (M\"uller et al.~2014, Tingay, Preston \& Jaucey~2001), TeV $\gamma$-ray emission can even originate at distances of the order of tens kpc. Therefore, it makes sense to investigate possible GeV-TeV $\gamma$-ray emission even from jets extending up to 30 kpc. 

The Cherenkov Telescope Array (CTA), with about three times better angular resolution and about an order of magnitude better sensitivity at $\sim$1~TeV than the present Cherenkov observatories (Acharya et al.~2013), will certainly provide results which will put new light on the importance of the high energy processes in the kpc scale jets. They will allow to constrain models for the jet propagation at such distances and also processes of electron acceleration. In fact, Angioni (2019) predicted that the CTA will be able to detect several new radio galaxies based on the extrapolation of the spectra from the GeV $\gamma$-ray energy range of radio galaxies reported recently in the Forth {\it Fermi}-LAT catalog (Abdollahi~et~al.~2019).
 
In the above discussed example calculations of the TeV $\gamma$-ray emission from the kpc scale jets, we do not take into account possible effects of $\gamma$-ray absorption in the starlight radiation of the host galaxy. In fact, the optical depths for such absorption process are predicted to be rather low (see e.g. Stawarz et al. 2006, Zacharias, Chen \& Wagner~2016).  

\begin{figure}
\vskip 9.truecm
\includegraphics{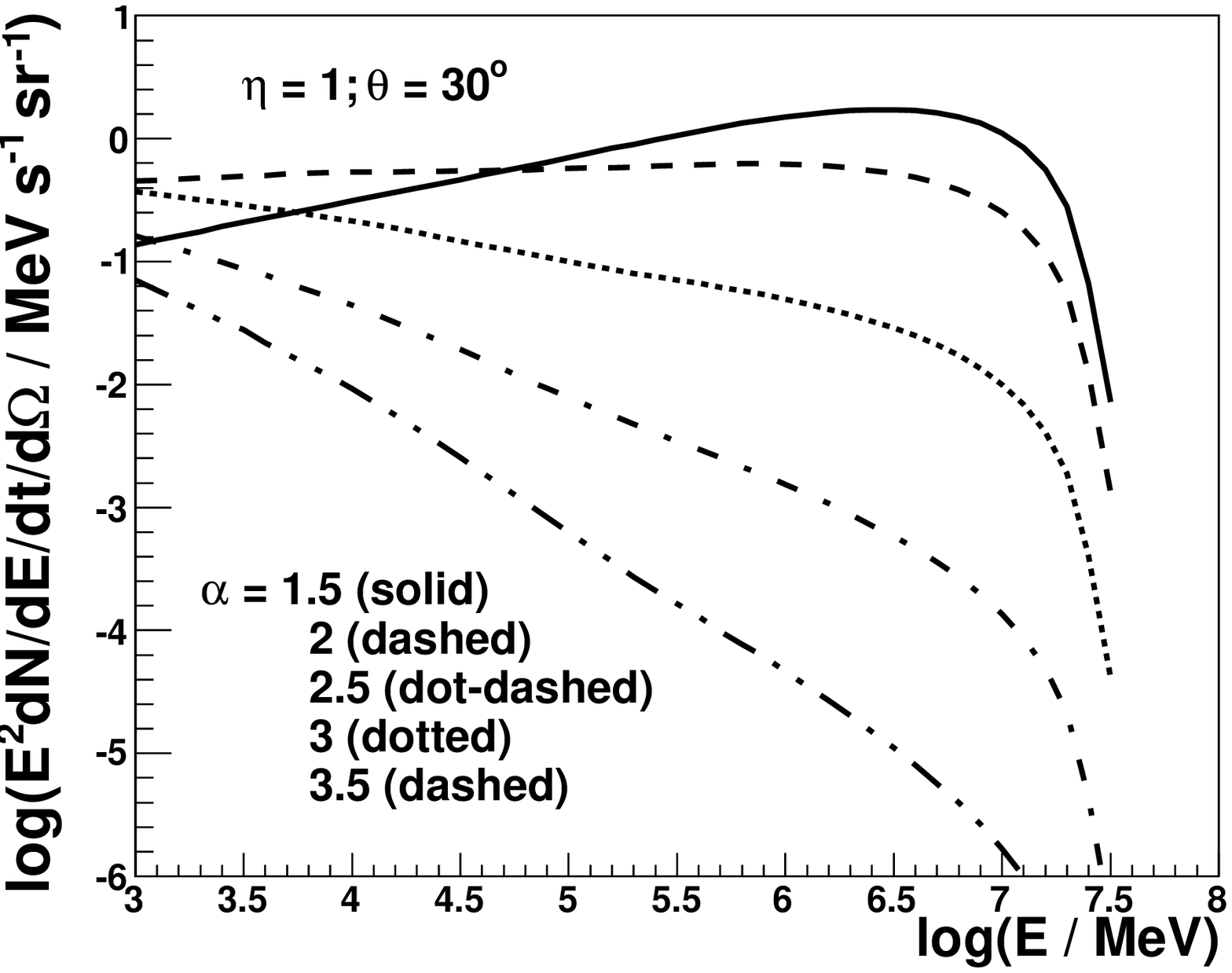}
\includegraphics{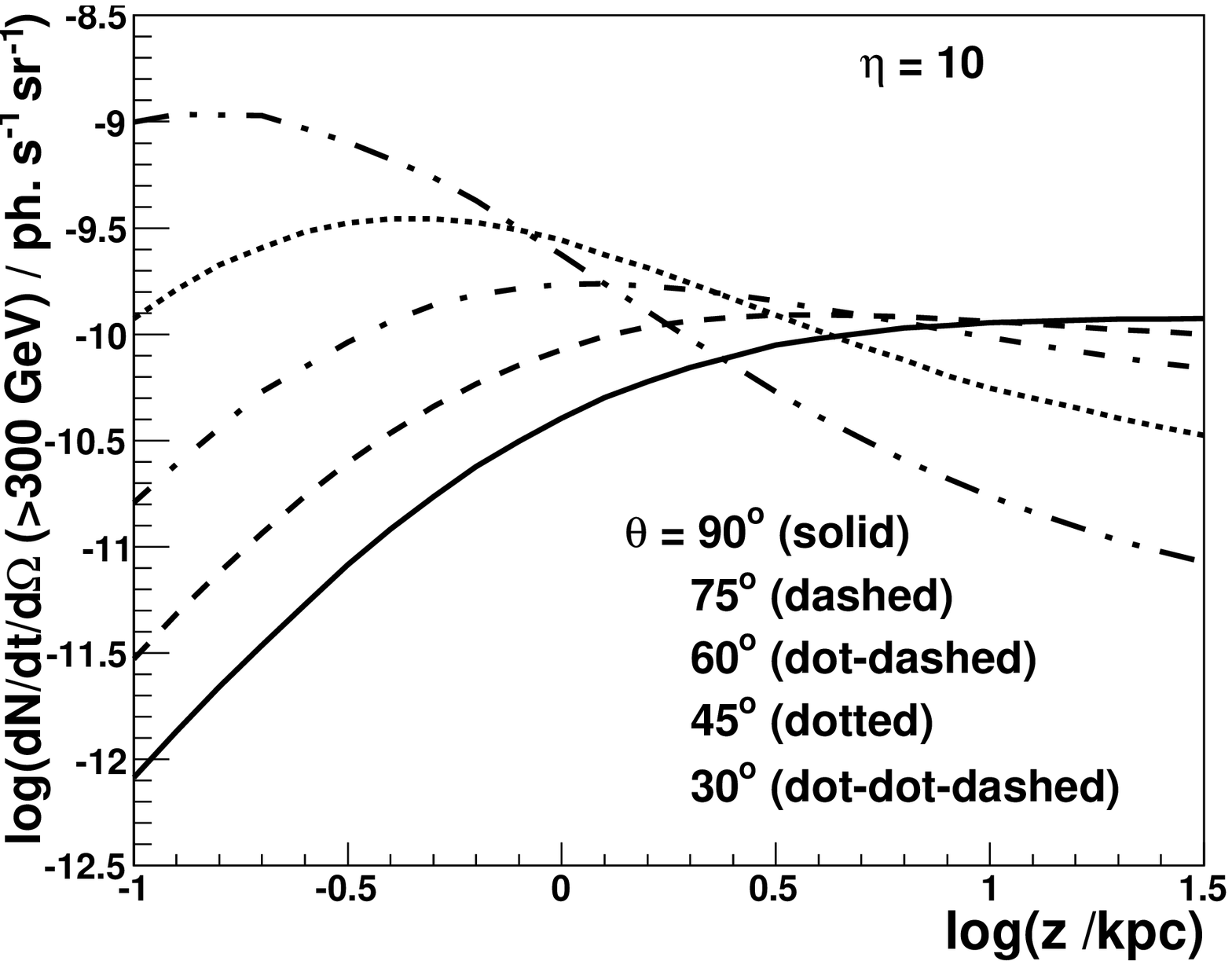}
\caption{Upper figure: The SEDs of $\gamma$-rays from the kpc scale jet, integrated over the range of distances between $z = 0.1$ kpc and 30 kpc (as in Fig.~13) but for different indexes of the electron spectrum: $\alpha = 1.5$ (solid), 2 (dashed), 2.5 (dot-dashed), 3 (dotted), and 3.5 (dot-dot-dashed). The observation angle is fixed on $\theta = 30^\circ$ and the parameter $\eta = 1$.
Bottom figure: The $\gamma$-ray flux ($>$300 GeV), as a function of the real distance from the base of the jet (for the parameters as in Fig.~8d) but for the index of the electron spectrum equal to $\alpha = 3$.}
\label{fig14}
\end{figure}

I would like to thank the Referee for many detailed comments and J. Sitarek for reading the manuscript and comments.
This work is supported by the grant through the Polish National Sci\-ence Cen\-ter No. UMO-2016/22/M/ST9/00583.
\vskip 1.5truecm
\leftline{\bf{\large{Bibliography}}}
\vskip 1truecm

\noindent
Abdalla, H., Abramowski, A., Aharonian, F. et al. 2018, A\&A, 619, 71

\noindent
Abdo, A.A., Ackermann, M., Ajello, M. et al. 2010, ApJ, 719, 1433

\noindent
Abdollahi, S. et al. 2019 ApJS, submitted (arXiv:1902.10045)

\noindent
Acciari, V.A., Ansoldi, S., Antonelli, L.A. et al. 2020 MNRAS, DOI:10.1093/mnras/staa014

\noindent
Acharya, B.S. et al. 2013, APh, 43, 3

\noindent
Aharonian, F.A., Atoyan A.M. 1981, Ap\&SS, 79, 321, 336

\noindent
Aharonian, F., Akhperjanian, A.G., Anton, G. et al. 2009, A\&A, 502, 749

\noindent
Angioni, R. 2019 APh, in press (arXiv:1910.00952)

\noindent
Banasi\'nski, P., Bednarek, W. 2018, ApJ, 864, 128

\noindent
Barthel, P.A. 1989, ApJ, 336, 606

\noindent
Bednarek, W. 2019, MNRAS, 483, 1003

\noindent
Blumenthal, G.R., Gould, R.J. 1970, Review of Modern Physics, 42, 237 

\noindent
Brown, A.M.  Bahm, C., Graham, J., Lacroix, T., Chadwick, P., Silk, J. 2017, PRD, 95, 063018

\noindent
Bradford, S., Nulsen, P.E.J., Kraft, R.P. et al. 2019, ApJ, 879, 8

\noindent
Bridle A H., Hough D. H., Lonsdale C. J., Burns J. 0., Laing R A,
1994, AJ, 108, 766

\noindent
Butuzova, M.S., Pushkarev, A.B. 2019, ApJ, 883, 131

\noindent
Chiaberge, M., Capetti, P.A., Celotti, A. 2001, MNRAS, 324, L33 

\noindent
Feigelson, E.D., Schreier, E.J., Delvaille, J.P., Giacconi, R., Grindlay, J.E., Lightman, A.P. 1981, ApJ, 251, 31

\noindent
Goodger, J.L. et al. 2010, ApJ, 708, 675

\noindent
Hardcastle, M. J.; Birkinshaw, M.; Worrall, D. M. 2001, MNRAS, 326, 1499 

\noindent
Hardcastle, M. J., Croston, J. H. 2011, MNRAS, 415, 133

\noindent
Hardcastle, M.J., Worrall, D.M., Kraft, R.P., Forman, W.R., Jones, C., Murray, S.S. 2003, ApJ, 593, 169

\noindent
Hardcastle, M.J., Kraft, R.P., Worrall, D.M. 2006, MNRAS, 368, L15

\noindent
Harris, G.L.H., Rejkuba, M., Harris, W.E. 2010, PASA, 4, 457

\noindent
Mannheim, K.; Biermann, P. L. 1992 A\&A 253, L21

\noindent
Maraschi, L., Ghisellini, G., Celotti, A. 1992 ApJ 397, L5

\noindent
Marconi, A., Schreier, E.J., Koekemoer, A., Capetti, A., Axon, D., Macchetto, D., Caon, N. 2000, ApJ, 528, 276

\noindent
Mirabel, I.F. et al. 1999, A\&A, 341, 667

\noindent
Moderski, R., Sikora, M., Coppi, P.S., Aharonian, F. 2005, MNRAS, 363, 954 

\noindent
M\"uller, C. et al. 2014, A\&A, 569, 115

\noindent
Perlman, E.S., Wilson, A.S. 2005, ApJ, 627, 140

\noindent
Pudritz, R.E., Hardcastle, M.J., Gabuzda, D.C. 2012, SSRv, 169, 27

\noindent
Sahakyan, N., Yang, R., Aharonian, F.A., Rieger, F.M. 2013, ApJ, 770, L6

\noindent
Sahakyan, N., V. Baghmanyan, V., Zargaryan, D. 2018, A\&A, 614, A6

\noindent
Sanchez, D. et al. for the HESS Collab. 2018, talk TeVPa in Berlin

\noindent
Sikora, M., Begelman, M.C., Rees, M.J. 1994 ApJ 421, 153

\noindent
Skibo, J.G., Dermer, C.D., Kinzer, R.L. 1994, ApJ, 426, L23

\noindent
Stawarz, \L, Sikora, M., Ostrowski, M. 2003, ApJ, 597, 186

\noindent
Stawarz L., Aharonian F., Wagner S., Ostrowski M., 2006, MNRAS, 371, 1705

\noindent
Tanada, K., Kataoka, J., Inoue, Y. 2019 ApJ, 878, 139

\noindent
Tingay, S.J., Preston, R.A., Jauncey, D.L. 2001, AJ, 122, 1697

\noindent
Urry C. M., Padovani P., 1995, PASP, 107, 803

\noindent
Wardle, J.F.C., Aaron, S.E. 1997, MNRAS, 286, 425

\noindent
Wykes, S., Hardcastle, M.J., Karakas, A.I., Vink J.S. 2015, MNRAS, 447, 1001

\noindent
Zacharias, M., Chen, X., Wagner, S.J. 2016, MNRAS, 465, 3767

\end{document}